\RequirePackage{fix-cm}
\documentclass[twocolumn]{svjour3}          
\smartqed  
\usepackage{graphicx}
\usepackage{mathptmx}      
\makeatletter
\expandafter\let\csname opt@amsmath.sty\endcsname\relax
\makeatother
\usepackage{amsmath,amssymb}
\usepackage{graphicx}
\usepackage{bbold}
\usepackage{color}
\newcommand{\rot}[1]{{\color{black} #1}}

\newcommand{\tl}{\hat{t}}
\newcommand{\tref}{t_\text{ref}}

\newcommand{\vth}{V_\text{T}}
\newcommand{\vreset}{V_\text{R}}

\newcommand{\taum}{\tau_\text{m}}
\newcommand{\taus}{\tau_\text{s}}
\newcommand{\tcor}{\tau_\text{cor}}
\newcommand{\isi}{P}

\newcommand{\isdef}{\stackrel{\text{\tiny def}}{=}}

\newcommand{\lrk}[1]{\left\langle #1 \right\rangle}
\newcommand{\lrrund}[1]{\left( #1 \right)}
\newcommand{\lreckig}[1]{\left[ #1 \right]}
\newcommand{\lrrundd}[1]{\bigl( #1 \bigr)}

\newcommand{\od}[2]{\frac{\mathrm{d}#1}{\mathrm{d}#2}}

\renewcommand{\vec}[1]{\mathbf{#1}}

\begin{document}

\title{Mapping Input Noise to Escape Noise in Integrate-and-fire neurons: A Level-Crossing Approach}


\titlerunning{From Input Noise to Escape Noise: A Level-Crossing Approach}        

\author{Tilo Schwalger}


\institute{T. Schwalger \at
Institute of Mathematics, Technical University Berlin, 10623 Berlin, Germany\\
              \email{schwalger@math.tu-berlin.de}           
           \and
           T. Schwalger \at
Bernstein Center for Computational Neuroscience Berlin, 10115 Berlin, Germany
}

\date{Received: date / Accepted: date}

\maketitle

\begin{abstract}
  Noise in spiking neurons is commonly modeled by a noisy input current or by generating output spikes stochastically with a voltage-dependent hazard rate (``escape noise''). While input noise lends itself to modeling biophysical noise processes, the phenomenological escape noise is mathematically more tractable. Using the level-crossing theory for differentiable Gaussian processes, we derive an approximate mapping between colored input noise and escape noise in leaky integrate-and-fire neurons. This mapping requires the first-passage-time (FPT) density of an overdamped Brownian particle driven by colored noise with respect to an arbitrarily moving boundary. Starting from the Wiener-Rice series for the FPT density, we \rot{apply} the second-order decoupling approximation of Stratonovich to the case of moving boundaries and derive a simplified hazard-rate representation that is local in time and numerically efficient. This simplification requires the calculation of the non-stationary auto-correlation function of the level-crossing process: For exponentially correlated input noise (Ornstein-Uhlenbeck process), we obtain an exact formula for the zero-lag auto-corre\-lation as a function of noise parameters, mean membrane potential and its speed, as well as an exponential approximation of the full auto-correlation function.  The theory well predicts the FPT and interspike interval densities as well as the population activities obtained from simulations with colored input noise and time-dependent stimulus or boundary. The agreement with simulations \rot{is strongly enhanced} across the sub- and suprathreshold firing regime \rot{compared to} a first-order decoupling approximation that neglects correlations between level crossings. \rot{The second-order approximation also improves upon a previously proposed theory in the subthreshold regime.} Depending on a simplicity-accuracy trade-off, all considered approximations represent useful mappings from colored input noise to escape noise, enabling progress in the theory of neuronal population dynamics.

\end{abstract}

\section{Introduction}
\label{intro}

Neurons in the brain must operate under highly non-stationary conditions. In fact, most behaviorally relevant sensory stimuli as well as internal signals are rarely constant in time but may change rapidly. In the presence of noise, such dynamic stimuli can be reliably encoded in the time-dependent population activity of a large population of spiking neurons \cite{GerKis14}. The time-dependent population activity also provides a concise, coarse-grained description of the collective dynamics of interacting spiking neurons.  Therefore, theories that predict the population activity in response to a time-dependent signal have been of fundamental interest in theoretical neuroscience \cite{Kni72,Ger00,AugLad17,SchDeg17}.

The population activity of noisy spiking neurons can be mathematically described by population density equations \cite{NykTra00,Chi17}. The form of the population density equation depends on the noise model. Two popular ways to model neuronal noise consist of  modeling noise either in the input or in the output of the neuron \cite{GerKis14}. In the first model class (\emph{input noise}), noise enters the dynamical equations of the membrane potential, currents or conductances leading to stochastic differential equations. If the noise is Gaussian white noise, the subthreshold dynamics becomes a diffusion process and the input noise is also called \emph{diffusive noise} \cite{Ger00}. The corresponding population density equation is a Fokker-Planck equation and the population activity can be obtained as the probability flux across the threshold \cite{AbbVre93,BruHak99,FouBru02,NykTra00,Ric08,AugLad17}. Models based on diffusive noise naturally appear as the result of modeling biophysical processes such as synaptic shot-noise or ion channel noise. In particular, a frequently considered source of noise is background synaptic input modeled as external Poisson processes \cite{Bru00,PotDie14}. The fluctuating part of this external shot noise leads, via a diffusion approximation \cite{GerKis14}, to Gaussian white noise driving the synaptic input current or conductance. Besides its biophysical interpretability, input noise has the advantage that it permits modeling both temporal \cite{FouBru02,SchDro15} and spatial \cite{LinDoi05} correlations of synaptic inputs and enables mean-field theories for recurrent networks of sparsely-connected integrate-and-fire neurons \cite{BruHak99,Bru00}.

In the second model class (called \emph{output noise} or \emph{escape noise} \cite{Ger00}), the dynamical equations for the state variables are deterministic while spikes (``output'') are generated stochastically through a hazard rate or conditional intensity \cite{Ger00,Pan04,TruEde05,PilShl08,PilLat08,NauGer12,BreSen13,GalLoe16,GerDeg17,RaaDit20}. This hazard rate depends on the state variables via a link function. For example, it may be given as $\hat\lambda(t)=\Psi(u(t),\tl(t))$, where $u(t)$ is the membrane potential and $\tl(t)$ is the last spike time of the neuron at time $t$. If the neuron model is a non-homogeneous renewal or quasi-renewal \cite{NauGer12} process, the corresponding population density equation is a renewal integral equation or, equivalently, a refractory density equation \cite{GerKis14,Ger00,NauGer12,ChiGra07,ChiGra08,DumHen16,SchChi19}. Although output noise is of phenomenological nature without a quantitative link to biophysical mechanisms, it has several advantages \cite{SchChi19} owing to its simpler mathematical tractability: First, the refractory density or integral equation admits an extension to finite numbers of neurons \cite{SchDeg17,SchChi19,SchGer20,SchLoe21_arxiv}. This extension allows to account for finite-size fluctuations of the population activity at the mesoscopic scale. Second, models with output noise provide analytical expressions for the likelihood function, and thus model parameters can be efficiently fitted to experimental data of single neuron recordings \cite{Pan04,TruEde05,PilShl08,GerDeg17,MenNau12,PozMen15,TeeIye18}. And third, the state space for models with output noise remains approximately one-dimensional even for multi-dimen\-sional conductance-based neuron models \cite{ChiGra07}. The one-dimensional description permits highly efficient numerical solutions, in contrast to Fokker-Planck equations \cite{ApfLy06}, which become intractable and computationally inefficient for several state variables.

In view of the wide use of biologically interpretable input noise and the mathematical advantages of output noise, an intriguing question is whether input noise can be approximately mapped to output noise, so as to take full advantage of both noise models. Mathematically, such a map requires the specification of the hazard rate $\hat\lambda(t)$ in terms of a link function $\Psi$, which depends on some dynamical variables and defines the escape-noise model. \rot{Unfortunately, a standard method to derive such a link function does not exist. To see this, let us consider the example of nonhomogeneous renewal processes as a popular class of neuron models.} In these models, the probability density $P(t|\tl)$ to fire the next spike at time $t$ given a spike at time $\tl$, $\tl<t$, does not depend on the state of the model before time $\tl$, i.e. the memory of renewal neurons only reaches back to its last spike. An important example of nonhomogeneous renewal models in neuroscience are one-dimensional integrate-and-fire neurons driven by white input noise \cite{GerKis14}. For this model class one can formally construct the hazard rate via the formula $\lambda(t|\tl)=P(t|\tl)/\bigl[1-\int_{\tl}^tP(s|\tl)\,ds\bigr]$ \cite{GerKis14}.
However, there are two obstacles: first, in order to apply this formula, the ``interspike interval (ISI) density''  $\isi(t|\tl)$ would be needed in analytical form for \emph{arbitrary}, time-dependent input currents $\{I(t')\}_{t'\in(\tl,t)}$ that occurred since the last spike. However, the calculation of the ISI density for time-dependent inputs is equivalent to a first-passage-time (FPT) problem with time-dependent parameters or boundary. The solution of this FPT problem requires the solution of  the Fokker-Planck equation with moving absorbing boundary, which is known to be a hard theoretical problem \cite{BulEls96,SchTal04,Lin04b}. Second, even if one succeeds to derive an approximate formula for the hazard rate $\lambda(t|\tl)$, it is still challenging to represent the hazard rate in the form of a link function $\Psi\bigl(u(t),\{z(t)\},\tl\bigr)$ that depends on some voltage-like variable $u(t)$, the last spike time $\tl$ and possibly further dynamical variables $\{z(t)\}$ \emph{locally in time} (as opposed to a ``non-local'' functional of $\{u(t'),z(t')\}_{t'\in(\tl,t)}$). 

Several theoretical studies have suggested approximate local hazard rates for leaky integrate-and fire (LIF) models driven by white \cite{PleGer2000,HerGer01,ChiGra07} or exponentially-correlated \cite{ChiGra08} Gaussian noise, or quasi-static (frozen) noise \cite{GoeDie08}. In this paper, we explore an alternative approach to the hazard rate and the first-passage-time density based on the theory of level crossings \cite{VSS2006b}. In Sec.~\ref{sec:model-1}, we introduce the LIF model with time-dependent driving and constant threshold and map this process an equivalent model with constant input and moving barrier. In Sec.~\ref{sec:model}, we consider the level crossing statistics with respect to this moving barrier and use the Wiener-Rice series and approximations thereof to provide formal expressions for the FPT density. These expressions form the starting point for deriving approximate hazard rates that are local in time. This derivation reveals some unexpected results concerning the correlations of level-crossings of Gaussian processes at small time lags (Sec.~\ref{sec:corr-funct-level}). Then, we turn to the LIF model and the problem of mapping input noise to escape noise (Secs.~\ref{sec:from-diffusive-noise})  and apply this map to predict the time-dependent population activity of LIF neurons with colored input noise (Sec.~\ref{sec:pop-act}). Each of the sections~\ref{sec:model}, ~\ref{sec:from-diffusive-noise} and \ref{sec:pop-act} closes with a comparison of the level-crossing theory with simulations and a previous theory by Chizhov and Graham \cite{ChiGra08}. 
Detailed derivations are provided in the Appendix.

\section{Leaky integrate-and-fire models and the associated first-passage-time problem}
\label{sec:model-1}

As a spiking neuron model with input noise, we consider  a leaky integrate-and-fire model  driven by synaptically filtered (``colored'') noise \cite{SchLSG08,GerKis14,SchDie15}. In this model, spikes are emitted whenever the membrane potential $V(t)$ reaches a threshold $\vth$. The subthreshold dynamics for $V<\vth$ can be written as
\begin{subequations}
  \label{eq:lif}
  \begin{align}
    \taum\dot V&=-V+\mu(t)+\eta(t),\\
    \taus\dot \eta&=-\eta+\sqrt{2\taus}\sigma_\eta\xi(t),
  \end{align}  
\end{subequations}
where $\taum$ is the membrane time constant and $\mu(t)=V_\text{rest}+RI(t)$ is the mean neuronal drive consisting of a constant resting potential $V_\text{rest}$ and a time-dependent input current $I(t)$ ($R$ denotes the membrane resistance). Furthermore,  $\eta(t)$ is a colored noise modeled as a one-dimensional Ornstein-Uhlenbeck process with correlation time $\taus$ and variance $\sigma_\eta^2$, \rot{and $\xi(t)$ is a zero-mean Gaussian white noise with auto-correlation function $\langle\xi(t)\xi(t')\rangle=\delta(t-t')$. The colored noise} captures the effect of various intrinsic and extrinsic noise sources, such as  fluctuations of synaptic background activity in vivo (shot noise due to random spike arrival from background neurons). After threshold crossing and  spike emission, $V(t)$ is reset to a reset potential $V_R$, $V_R<V_T$, and the subthreshold dynamics Eq.~\eqref{eq:lif} resumes after an absolute refractory period of length $\tref$ following the reset.

We are seeking a corresponding spiking neuron model with escape-noise \cite{Ger00} given by a hazard rate (conditional intensity) of the form $\Psi\bigl(u(t),\dot u(t),\{z_i(t)\},t-\tl\bigr)$. Here, $u(t)$ is a membrane-potential variable that obeys the noiseless membrane dynamics of the LIF model  between spikes:
\begin{subequations}
  \label{eq:escape-noise-model}
  \begin{equation}
    \label{eq:u-dyn}
    \taum\dot u=-u+\mu(t). 
 \end{equation}
 Furthermore, we allow an explicit dependence on the speed of the membrane potential $\dot u(t)$ (in accordance with previous studies \cite{PleGer2000,HerGer01,ChiGra07,GoeDie08}), the time since the last spike $t-\tl$, and possibly further auxiliary variables $\{z_i\}$ whose dynamics between spikes is given by ordinary differential equations. Given these variables at time $t$, a spike is fired \rot{independently} in the next time step  with probability
\begin{multline}
  \text{Pr}(\text{spike in $(t,t+dt)$}|u(t),\dot u(t),\{z_i(t)\},t-\tl)\\
  =\Psi\bigl(u(t),\dot u(t),\rot{\{z_i(t)\}},t-\tl\bigr)dt
  \label{eq:hazard-escape}
\end{multline}
\end{subequations}
where $dt$ is a small step size. \rot{This probabilistic firing rule is the counterpart of the firing rule with a hard threshold  in the LIF model with input noise.}
After a spike, $u(t)$ is reset to $\vreset$  and the auxiliary variables $\{z_i\}$ are also reset to some suitable fixed reset value. During an absolute refractory period of length $\tref$, the variables are clamped to their reset values and the hazard rate is set to zero. Because all memory is erased upon resetting, the escape-noise model  is a non-homogeneous renewal process . 

The main goal is to map the model with colored input noise, Eq.~\eqref{eq:lif} to the model with escape noise, Eq.~\eqref{eq:escape-noise-model}. Strictly speaking, mapping the two models is an ill-posed problem because the model with input noise is a non-renewal process, whereas the escape-noise model is a (non-homogeneous) renewal process. In fact, the temporal correlations of the colored noise in Eq.~\eqref{eq:lif} introduces memory that is not erased upon spiking. This memory leads to correlations between interspike intervals (ISIs) \cite{Lin04,SchLSG08,SchDro15}. However, if the correlation time  $\taus$ of the colored noise is much smaller than the mean interspike interval, these correlations will be small and the model with input noise can be regarded as approximately renewal. In this case, it is sufficient to match the ISI densities of the two models in order to obtain an approximate mapping. Therefore, our goal of mapping the two models can be phrased more modestly as follows: Can we find a link function $\Psi$ of the escape-noise model such that for an arbitrary given stimulus $\mu(t)$ the time-dependent ISI densities $P(t|\tl)$ of the two models approximately match for all $t$ and $\tl<t$? 
We emphasize that this definition of the mapping rests on the assumption of sufficiently small correlation times of the colored input noise. Biologically, this assumption seems to be reasonable given that typical time scales of excitatory and inhibitory postsynaptic currents are often only on the order of a few milliseconds \cite{GerKis14}. 

To derive the link function $\Psi$ that maps input to output noise, one needs to solve a first-passage-time (FPT) problem: As mentioned in the introduction, the hazard rate can be obtained from the ISI density of the model with input noise, Eq.~\eqref{eq:lif}. In this model, the interspike interval is determined by the ``first-passage time'' that is needed for the membrane potential to travel from the reset potential to the threshold. Thus, the ISI density $P(t|\tl)$ is equivalent to the FPT density (apart from a time shift  due to the deterministic absolute refractory period). \rot{To compute the FPT density, one needs to choose suitable initial conditions for} the colored noise $\eta(t)$. The ISI starting at the last spike time $\tl$ is composed of the initial absolute refractory period of length $\tref$ and the stochastic FPT $t^*$. We thus need the initial value $\eta(\tl+\tref)$ of the noise at the starting time $\tl+\tref$ of the stochastic motion. At the firing time $\tl$, the distribution of the noise $p_\text{fire}(\eta,\tl)$ is biased towards positive values of $\eta$ \cite{Lin04,SchLSG08,Sch13,SchDro15}, in contrast to the stationary distribution  $p_\text{st}(\eta)$ of the Ornstein-Uhlenbeck noise, which has zero mean. During the absolute refractory period, the noise distribution relaxes towards the stationary distribution. \rot{Even though the noise at time $\tl+\tref$ may not be fully stationary yet, it is reasonable to assume stationary initial conditions, where $\eta(\tl+\tref)\sim \mathcal{N}(0,\sigma_\eta^2)$ is drawn from a normal distribution with variance $\sigma_\eta^2$. This initial condition is justified because the noise correlation time $\taus$ has been assumed to be much smaller than the mean ISI; hence, we do not expect that the precise shape of the initial noise distribution has a significant effect on the FPT density.}

\begin{figure*}[t]
  \centering
  \includegraphics{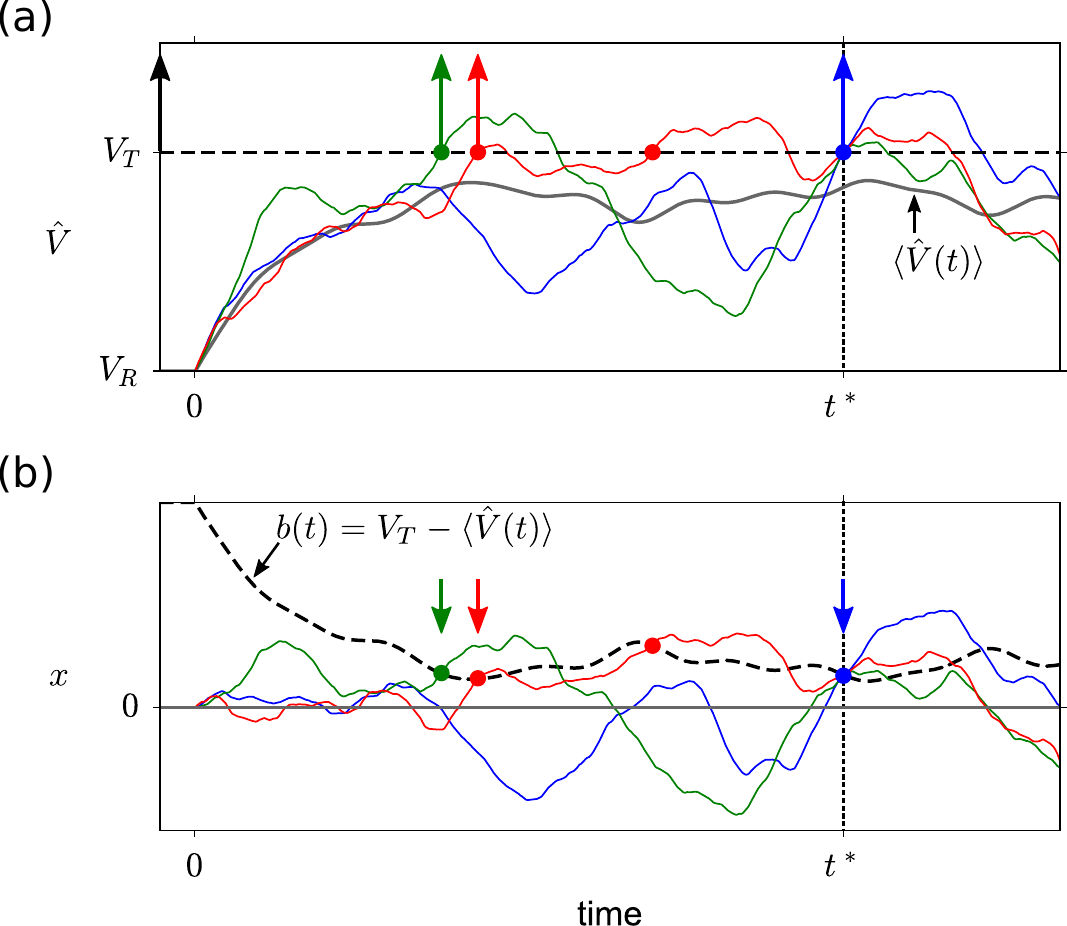}
  \caption{First-passage time of an integrate-and-fire neuron
    model and an equivalent model with moving boundary. (a) At time $t=0$, different realizations of the \rot{non-resetting} membrane
    potential \rot{$\hat V(t)$} (colored thin lines) are released from the reset
    potential $V_R$. The \rot{non-resetting}
    membrane potential follows a Gaussian process with time-dependent
    mean \rot{$\langle \hat V(t)\rangle$} (gray thick line). Shown are three
    realizations (green, red, blue lines) that have an identical threshold crossing at time
    $t=t^*$ (blue circle), \rot{which is not necessarily the first crossing (indicated by an arrow)}. (b) Transformation to an equivalent time-homogeneous process $x(t)$ with moving boundary $b(t)$, in which the positions of threshold crossings are preserved. Parameters: $\taus=4$~ms, $\taum=10$~ms, $\sigma_V:=\sigma_x(\infty)=0.25(\vth-\vreset)$.}
  \label{fig:levelcross-illu}
\end{figure*}

Because in the following we focus on the FPT starting at $\tl+\tref$, we will conveniently choose the time origin such that $\tl+\tref=0$. \rot{Furthermore, since we are only interested in the \emph{first} threshold crossing after time $t=0$, we can omit the voltage resetting for $t>0$ without changing the FPT statistics. The resulting non-resetting process $\hat V(t)$ is the freely evolving solution of Eq.~\eqref{eq:lif} without reset and} with initial conditions $\rot{\hat V}(0)=\vreset$, $\eta(0)\sim \mathcal{N}(0,\sigma_\eta^2)$ (Fig.~\ref{fig:levelcross-illu}a). \rot{This non-resetting process will be useful for the level-crossing approach below.}

For mathematical convenience, we will now reformulate the FPT problem in terms of a time-homogeneous process $x(t)$ and a moving boundary $b(t)$, so as to eliminate the time-dependent parameter $\mu(t)$ in Eq.~\eqref{eq:lif} (Fig.~\ref{fig:levelcross-illu}b). This is achieved by subtracting the mean \rot{non-resetting} membrane potential $\langle \rot{\hat V}(t)\rangle=u(t)$:
\begin{align}
  x(t)&=\rot{\hat V}(t)-u(t)\\
    \label{eq:b-u}
  b(t)&=V_T-u(t),
\end{align}
where $u(t)$ is given by Eq.~\eqref{eq:u-dyn} with initial condition $u(0)=\vreset$. Furthermore, setting $y=\eta/\taum$, $\gamma=1/\taum$, $\tau_y=\taus$ and $D=\taus\sigma_\eta^2/\taum^2$, we find the Langevin equation
\begin{subequations}
  \label{eq:25}
\begin{align}
\label{eq:2}
  \dot x&=-\gamma x+y\\
  \tau_y\dot y&=-y+\sqrt{2D}\xi(t)
\end{align}  
\end{subequations}
with initial conditions
\begin{equation}
  \label{eq:init-cond}
  x(0)=0,\qquad y(0)\sim \mathcal{N}(0,\sigma_y^2)
\end{equation}
The dynamics of $x(t)$ can be interpreted as an overdamped motion of Brownian particle in a parabolic potential subject to a colored noise $y(t)$ (Ornstein-Uhlenbeck process). Here, $D$ and $\tau_y$ are intensity and  the correlation time of the noise, respectively, and $\gamma$  is the friction coefficient. As before, $\xi(t)$ is a zero-mean Gaussian white noise. At time $t=0$, the random initial condition for the colored noise $y$ corresponds to a stationary Gaussian distribution with mean zero and variance $\sigma_y^2=D/\tau_y$. By construction, the domain of the particle is bounded from above by the time-dependent boundary $b(t)$, where $b(0)>0$ and $b(t)$ is a differentiable function of time. The FPT $t^*$ is defined as the time when $x(t)$ exits the domain, i.e. when it reaches the boundary,  for the first time. The FPT density will be denoted by $P(t)$, i.e. $P(t)dt=\text{Prob}(t^*\in[t,t+dt))$ for an infinitesimal time interval of length $dt$. We emphasize again that the FPT density of the Brownian particle $x(t)$ with moving boundary $b(t)$ is the same as the FPT density of the membrane potential $V(t)$ with respect to the constant threshold $\vth$. 

Beyond neuroscience, the escape of the doubly low-pass filtered process, Eq.~\eqref{eq:25}, from a domain with  moving boundary $b(t)$ may serve as a simple archetypal model for non-stationary FPT problems. One prominent example are reaction times of bimolecular chemical reactions \cite{HanTal90}. If $x(t)$ is interpreted as a reaction coordinate and the domain $x<b(t)$ corresponds to the reactant state, the boundary $b(t)$ can be interpreted as a time-dependent energy barrier that needs to be surpassed to reach the product state. Accordingly, the first-passage time can be interpreted as the reaction time.

 \section{Level-crossing theory for a moving barrier}
\label{sec:model}

\subsection{Hazard-rate representation of first-passage-time density}
\label{sec:haz-app}

To find approximations to the FPT density \rot{from approximate hazard rates}, we use concepts from renewal theory, especially the notion of hazard rate and survival probability \cite{Cox62}. \rot{Because the process Eq.~\eqref{eq:25} starts at time $0$,} the hazard rate $\lambda(t)$  is defined \rot{here} as the conditional probability per small time interval $dt$ to find a boundary crossing in the interval $(t,t+dt)$ {\em given} the absence of crossings \rot{in the interval $(0,t)$}. On the other hand, the survival probability $S(t)$ is defined as the probability of an absence of crossings \rot{in $(0,t)$. The two definitions imply that $S(t+dt)=S(t)(1-\lambda(t)dt)$, hence $dS(t)/dt=-\lambda(t)S(t)$. Because} the survival probability is unity at time $t=0$, we thus obtain $S(t)=\exp\lrrund{-\int_{0}^t\lambda(s)\,ds}$  for $t>0$. The probability to find the {\em first} crossing \rot{after time $0$} in the interval $(t,t+dt)$ is equal to the probability to find a crossing in $(t,t+dt)$ {\em and} to have no crossings in $(0,t)$. Hence, the FPT density is given by the product $P(t)=\lambda(t)S(t)$, or
\begin{equation}
  \label{eq:f-lam}
  P(t)=\lambda(t)\exp\lrrund{-\int_{0}^t\lambda(s)\,ds}.
\end{equation}
Given the hazard rate $\lambda(t)$ for $t>0$, Eq.~\eqref{eq:f-lam} provides a simple formula for the FPT density. An advantage of this representation is that the exponential factor can be turned into a first-order differential equation,
\begin{equation}
  \label{eq:S-ode}
  P(t)=\lambda(t)S(t),\qquad \od{S}{t}=-\lambda(t)S(t),\qquad S(0)=1.
\end{equation}
Thus, if the hazard rate $\lambda(t)$ can be efficiently computed for $t>0$, this representation permits an efficient numerical integration  of the first-passage-time density forward in time.
Therefore, the main strategy in this paper is to derive computationally efficient approximations for the hazard rate.  

In general, the calculation of the hazard rate is as difficult as the calculation of the FPT density itself. However, finding approximations for $\lambda(t)$ has several advantages over direct approximations of $P(t)$. Firstly, as a probability density, $P(t)$ must satisfy the normalization to unity. Thus, the value of the FPT density at different times cannot be calculated independently. In particular, the value of $P(t)$ strongly depends on the values for $t'\in(0,t)$. By contrast, $\lambda(t)$ is not a probability density and can thus, in principle, be  arbitrary as long as it is non-negative and $S(t)=\exp\lrrund{-\int_{0}^t\lambda(s)\,ds}$ converges to zero as $t\rightarrow\infty$. Thus, if we are able to find any approximation for $\lambda(t)$, the normalization of $P(t)$ is guaranteed by Eq.~\eqref{eq:f-lam}.

Secondly, the character of the hazard rate is more local in time than the FPT density, and thus, we expect more efficient approximations for the hazard rate. The non-local character of $P(t)$ has been already mentioned above. Moreover, the non-locality becomes particularly evident by the integral in Eq.~\eqref{eq:f-lam}, which accumulates the history of hazard rates. The exponential factor $S(t)$ shaped by this integral thus contributes a trivial history-dependence of the FPT density $P(t)$, which is present already for time-homogeneous processes. By contrast, this trivial history-dependence is divided out in the hazard rate $\lambda(t)=P(t)/S(t)$. The remaining time-dependence of the hazard rate singles out  effects of non-stationarity and explicit time-dependence of the system, which can be captured by local variables. Thirdly, because of the locality in time, time-dependent rates are interesting in its own right as they are often the natural choice to model escape processes in terms of a Markovian dynamics and master equations.

From the above considerations it becomes clear that the hazard rate representation, Eq.~\eqref{eq:f-lam}, is only useful if we succeed to derive approximations for $\lambda(t)$ that are local in time. This means that we are seeking an approximation of the hazard rate in the form
\begin{equation}
  \label{eq:haz-local}
  \lambda(t)\approx\Phi\bigl(b(t),\dot b(t),\dotsc, \{z_i(t)\},t\bigr),
\end{equation}
which may depend on time explicitly and through a few variables such as the value and its derivative of the time-dependent boundary, $b(t)$ and $\dot b(t)$, respectively, and possibly through a few auxiliary variables $z_i(t)$ that obey simple ordinary differential equations.
\rot{Note that we use the notations $\Phi$ for the boundary-dependent hazard rate of the model Eq.~\eqref{eq:25} and $\Psi$ for the voltage-dependent hazard rate of the model Eq.~\eqref{eq:hazard-escape}. The two functions are related in a simple way, see Sec.~\ref{sec:link-func}.}

\subsection{Wiener-rice series}
\label{sec:wiener-rice}

Our approach to tackle the time-dependent FPT problem is to employ the level-crossing statistics of a Gaussian process  \cite{Ric45,RicSat83,VSS2006b,BraThu17,AzaWsc09}. To this end, let us consider the sub-set of all realizations of $x(t)$ that cross the barrier $b(t)$ from below in the time interval $(t^*,t^*+\Delta t)$, a so-called ``up-crossing'' (Fig.~\ref{fig:levelcross-illu}b). The up-crossing at time $t^*$ is not necessarily the first one but could be the second, third (and so on) up-crossing (e.g. green and red lines in Fig.~\ref{fig:levelcross-illu}b). To compute the density of the first up-crossing, one can make use of the statistics of repeated up-crossing events. These events form a point process in \rot{the} time interval $[0,t^*]$
\begin{equation}
  \label{eq:spike-train}
  s(t)=\sum_{i=1}^{N(t^*)}\delta(t-\hat{t}_i), 
\end{equation}
where $N(t^*)$ denotes the (random) number of up-crossings in that interval, \rot{$\{\hat{t}_i\}_{i=1,\dotsc,N(t^*)}$ are the up-crossing times} and $\delta(\cdot)$ is the Dirac $\delta$-function. The statistics of the point process can be fully described by the set of moment functions $f_k(t_1,\dotsc,t_k)=\langle s(t_1)\dotsb s(t_k)\rangle$, for $k=1,2,\dotsc$ and non-coinciding time arguments $t_i$ \cite{Str67I,van92}. The moment functions can be interpreted such that for a small time step $\Delta t$ the quantity $f_k(t_1,\dotsc,t_k)\Delta t^k$ yields the probability to find an up-crossing events in each of the non-overlapping intervals $(t_1,t_1+\Delta t)$, ..., $(t_k,t_k+\Delta t)$. For instance, $f_1(t)$ yields the rate of up-crossings at time $t$, and $f_2(t_1,t_2)/f(t_1)$ is the conditional rate of an upcrossing at time $t_2$ given an up-crossing at time $t_1$.  For level-crossings of Gaussian processes, the distribution functions $f_k$ can be calculated explicitly, both for stationary and non-stationary processes (see appendix, Sec.~\ref{sec:fk}).

The distribution functions $f_k$ allow for an exact series expression of the FPT density, sometimes called Wiener-Rice series \cite{VSS2006b,BraThu17}:
\begin{equation}
  \label{eq:wiener-rice}
  P(t)=\sum_{k=0}^\infty \frac{(-1)^k}{k!}\int_{0}^tdt_1\dotsb dt_k\,f_{k+1}(t_1,\dotsc,t_k,t)
\end{equation}
A detailed explanation of this formula is given in reference \cite{VSS2006b}. In brief, it counts -- for a large ensemble of trajectories -- the number of those trajectories that have a crossing in $[t,t+dt)$ but no crossing in $(0,t)$. Starting with the fraction $f_1(t)dt$ of all trajectories that cross the boundary at time $t$ ($k=0$ term), the fraction with no previous crossing can be computed by subtracting those trajectories that crossed the boundary before time $t$. The second term $\int_0^tf_2(t_1,t)dt_1$ in Eq.~\eqref{eq:wiener-rice} accounts for these trajectories but overestimates their number because some trajectories are counted multiply. This corresponds to trajectories that cross the boundary more than once before time $t$ (e.g. red line in Fig.~\ref{fig:levelcross-illu}).  To correct for the excessive subtraction, one needs to add the fraction of trajectories with two or more crossings before $t$. This is taken into account by the third term $\frac{1}{2}\int_0^t\int_0^tf_3(t_1,t_2,t)dt_1dt_2$ which computes the mean number of crossing pairs $\{\hat{t}_1,\hat{t}_2\}$ per trajectory (e.g. in Fig.~\ref{fig:levelcross-illu}, the blue and green curve contributes zero and the red curve contributes one such pair; the factor $\frac{1}{2}$ accounts for permutations of $\hat{t}_1$ and $\hat{t}_2$). Again, this term overestimates the fraction of trajectories with double crossing events because trajectories with more than two crossings are multiply counted (e.g. a trajectory with three crossings gives rise to three pairs $\{\hat{t}_1,\hat{t}_2\}$, $\{\hat{t}_1,\hat{t}_3\}$, $\{\hat{t}_2,\hat{t}_3\}$). Continuing this correction procedure for trajectories with arbitrary number of crossings leads to the infinite series expression Eq.~\eqref{eq:wiener-rice}.

An alternative statistical description of the point process $s(t)$ is given by the $k$-th order cumulant functions $g_k(t_1,\dotsc,t_k)$ (see \cite{Str67I,van92} and Sec.~\ref{sec:gen-surv}), which remove the dependence on lower-order moment functions: for instance, $g_1(t)=f_1(t)$ and $g_2(t_1,t_2)=f_2(t_1,t_2)-f_1(t_1)f_1(t_2)$. The probability to find no event in the interval $(0,t)$ (i.e. the survival probability) is related to the cumulant functions by \cite{Str67I,van92}
\begin{equation}
  \label{eq:S-full}
  S(t)=\exp\left(\sum_{k=1}^\infty\frac{(-1)^k}{k!}\int_{0}^t\mathrm{d}t_1\dotsb\mathrm{d}t_k\,g_k(t_1,\dotsc,t_k)\right).
\end{equation}
From this expression, the Wiener-Rice series for the FPT density,
Eq.~\eqref{eq:wiener-rice} is recovered by
$P(t)=-dS(t)/dt$. Similarly, the hazard rate can be obtained by
$\lambda(t)=-d(\ln S)/dt$. As infinite series expressions, Eq.~\eqref{eq:wiener-rice} and Eq.~\eqref{eq:S-full} are of no practical use for direct computations of the FPT density. However, these formal expressions are used as a starting point for further approximations.

\subsection{Decoupling approximations}
\label{sec:decoupli}

The series expression for the survival probability, Eq.~\eqref{eq:S-full}, simplifies considerably if higher-order cumulant functions $g_k$ are approximated in terms of lower-order cumulant functions, thereby neglecting higher-order dependencies between up-crossings. In this section, we review two approximations based on such a decoupling of (temporal) interactions between events \cite{Str67I}: a first-order decoupling approximation, where all up-crossing events are assumed to be independent, and a second-order decoupling approximation, in which higher-order interactions are modeled in terms of pairwise interactions. While the first-order approximation readily results in local hazard rates, the more accurate pairwise interaction approximation is highly non-local and therefore not useful for practical calculations. However, as we shall show in Sec.~\ref{sec:local-hazard-funct}, the pairwise interaction model can be used as a starting point for deriving an efficient local approximation of the hazard rate \rot{(second-order decoupling approximation)} that accounts for higher-order interactions between up-crossings.

\subsubsection{Independent upcrossings}
\label{sec:poisson-upcr}

If the correlation time of the process $x(t)$ is much smaller than the (typical) intervals between
upcrossings, up-crossing events can be regarded as independent, i.e. the series of up-crossing events is an inhomogeneous Poisson process with rate $f_1(t)$. Mathematically, this corresponds to neglecting higher-order cumulants except for the first one: $g_1(t)=f_1(t)$ and $g_k\approx 0$ for all $k\ge 2$ \cite{Str67I}. In this case, Eq.~\eqref{eq:S-full} reduces to $S(t)=\exp\left(-\int_{0}^{t}f_1(\tau)\,d\tau\right)$, and hence the FPT density reads
\begin{equation}
  \label{eq:58}
  P(t)\approx f_1(t)\exp\left\{-\int_{0}^{t}f_1(\tau)\,d\tau\right\}.
\end{equation}
From this expression, we see that the hazard rate is simply given by the upcrossing rate of the freely evolving process $x(t)$: $\lambda(t)\approx f_1(t)$.   The upcrossing rate $f_1(t)$ can be calculated analytically in terms of the current value of the boundary $b(t)$ and its derivative $\dot b(t)$ (see Appendix~\ref{sec:fk} and \ref{sec:upross-rate-f_1}). The result is \rot{the first-order decoupling approximation:}
\begin{multline}
  \lambda(t)\approx f_1(t)=\Phi_1\bigl(b(t),\dot b(t),t\bigr)\\
  :=\frac{\sqrt{\sigma_x^2\sigma_y^2-\sigma_{xy}^2}}{2\pi\sigma_x^2}H\left(\frac{(\gamma\sigma_x^2-\sigma_{xy})b+\sigma_x^2\dot b}{\sqrt{2(\sigma_x^2\sigma_y^2-\sigma_{xy}^2)}\sigma_x}\right)e^{-B(b,\dot b,t)},
  \label{eq:n1}
\end{multline}
where $H(x)=1-\sqrt{\pi}xe^{x^2}\text{erfc}(x)$ and
\begin{align}
  \label{eq:Bfunc}
  B(b,\dot b,t)&=\frac{(\gamma^{2}\sigma_x^2-2\gamma\sigma_{xy}+\sigma_y^2)b^2+2(\gamma\sigma_x^2-\sigma_{xy})b\dot b+\sigma_x^2\dot b^2}{2(\sigma_x^2\sigma_y^2-\sigma_{xy}^2)}.
\end{align}
In these equations, the time-dependent moments $\sigma_{xy}(t)=\langle x(t)y(t)\rangle$ and $\sigma_x^2(t)=\langle x^2(t)\rangle$ are given by
\begin{subequations}
  \label{eq:sigmas-main}
\begin{align}
  \sigma_{xy}(t)&=\tilde{\tau}\sigma_y^2\left(1-e^{-t/\tilde{\tau}}\right),\\
  \sigma_x^2(t)&=\frac{\tilde{\tau}\sigma_y^2}{\gamma}\left(1-e^{-2\gamma t}\right)+\frac{2\tilde{\tau}\sigma_y^2}{2\gamma-\tilde{\tau}^{-1}}\left(e^{-2\gamma t}-e^{-\frac{t}{\tilde{\tau}}}\right)
\end{align}  
\end{subequations}
with $\sigma_y^2=D/\tau_y$ and $\tilde{\tau}^{-1}=\gamma+\tau_y^{-1}$ (see Sec.~\ref{sec:momen}, esp. Eq.~\eqref{eq:sigmadgl} for a numerically stable ODE representation of the moments).

Similar expressions for the level-crossing density in the time-inhomogeneous case have been derived in previous studies \cite{RicSat83,Bad11}.   

\subsubsection{Upcrossings correlated in pairs}

If the average time between  upcrossings $1/f_1(t)$ is on the order of or smaller than the correlation time of $x(t)$ given by $\tau_{cor}=\gamma^{-1}+\tau_y$, upcrossing events cannot be regarded as being independent anymore. To account for correlations between upcrossings, we follow a decoupling approximation (DA) of higher-order correlation functions $g_k(t_1,\dotsc,t_k)$, $k\ge 3$, proposed by Stratonovich \cite{Str67I,Str67II}. This approximation assumes that higher-order correlations are governed by the same time scales as pair-wise correlations and can therefore be expressed in terms of the first two correlation functions $f_1(t)$ and $g_2(t_1,t_2)$.  Specifically, correlation functions with $k\ge 2$ are approximated by the ansatz \rot{\cite{Str67I,Str67II}}
\begin{equation}
  \label{eq:DECOUPL-strat}
  g_k(t_1,\dotsc,t_k)=(k-1)!f_1(t_1)\dotsb f_1(t_k)\{R(t_2,t_1)\dotsb R(t_k,t_1)\}_{sym}.
\end{equation}
Here, the function $R(t,t')$ describes the pairwise interactions between events at time
$t$ and $t'$, and $\{\dotsb\}_{sym}$ denotes the operation of symmetrization (i.e. the
arithmetic mean of all permutations of the time arguments). \rot{As suggested in \cite{Str67I,Str67II}}, we
choose $R(t,t')$ as the normalized {\em auto-correlation function}
\begin{equation}
  \label{eq:60}
  R(t,t')=\frac{f_2(t,t')}{f_1(t)f_1(t')}-1,
\end{equation}
which makes the ansatz Eq.~\eqref{eq:DECOUPL-strat} exact for $k=2$. Note that compared to
\cite{Str67I,Str67II}, we use an opposite sign in the definition of $R$ for mathematical convenience. The auto-correlation function $R(t,t')$ can be interpreted as the conditional probability density of an event at time $t'$ given an event at time $t$ normalized by the unconditional probability density $f_1(t')$ and shifted by the mean such that $R(t,t')=0$ if events at time $t$ and $t'$ are independent. For stationary point processes, $R(t,t')=R(|t-t'|)$ only depends on the time difference. In analogy to the common use for spatial point processes, $R(t-t')$ will be called {\em pair correlation function} in this case.

We expect the following behavior of the auto-correlation function: firstly, if events are far apart, $|t-t'|\gg\tcor$, they occur independently, hence $f_2(t,t')\approx f_1(t)f_1(t')$. This implies a
vanishing auto-correlation function $R(t,t')\approx 0$.  Secondly, the behavior when $t$ and $t'$ are close depends on the correlations between events: if close events occur independently as in the case of an \rot{inhomogeneous} Poisson process, $R(t,t')$ vanishes. In contrast, a
positive pair correlation function $R(t,t')>0$ at small time lag indicates that events are {\em attractive} and tend to cluster. Conversely, for a negative pair correlation function $R(t,t')<0$ at small time lag, events are {\em repulsive}, i.e. the occurrence of close events is less frequent than expected for a Poisson process. In particular, if a point process exhibits a refractory period after each event (``hardcore interaction''), we find that $f_2(t,t')=0$ and hence  $R(t,t')=-1$ if $t$ and $t'$ fall within a refractory period. Similarly, non-approaching random points \cite{Str67I} are characterized by $R(t,t)=-1$ in the limit of vanishing time lag. Interestingly, it has been assumed by some authors that level crossings of differentiable processes are non-approaching events with $R(t,t)=-1$ \cite{VSS2006b,PueWol16}. In Sec.~\ref{sec:corr-funct-level} we shall investigate this assumption in more detail.   

\rot{While the decoupling approximation (DA), Eq.~\eqref{eq:DECOUPL-strat}, is exact for $k=2$ by construction, it must be considered as a physically motivated, heuristic ansatz for $k\ge 3$, which in general is not expected to be exact. Nevertheless, the ansatz and the above-described behavior of $R(t,t')$ ensure some important properties of the higher-order correlation functions $g_k$: first, the DA is exact for an inhomogeneous Poisson process because in this case $R(t,t')\equiv 0$ and thus  Eq.~\eqref{eq:DECOUPL-strat} recovers the expected result $g_k\equiv 0$ for all $k\ge 2$. Second, $g_k$ does not depend on the order of the time arguments because of the symmetrization operation in Eq.~\eqref{eq:DECOUPL-strat}. Third, $g_k(t_1,\dotsc,t_k)\approx 0$ if the time difference of two arguments is much larger than $\tcor$ because their pair correlation vanishes. And forth, it is known that for a system of non-approaching random points $g_k(t,\dotsc,t)=(-1)^k(k-1)!f_1^k(t)$  \cite{Str67II}, which is consistent with Eq.~\eqref{eq:DECOUPL-strat} and $R(t,t)=-1$.}

Substituting the DA, Eq.~\eqref{eq:DECOUPL-strat}, into the general expression for the survival probability,
Eq.~\eqref{eq:S-full}, yields \cite{Str67I,Str67II,VSS2006b,MeeAlb21_arxiv}
\begin{equation}
  \label{eq:22}
  S(t)\approx\exp\left\{-\int_{0}^{t}f_1(\tau)\frac{\ln\left[1+q(t,\tau)\right]}{q(t,\tau)}\,d\tau\right\},
\end{equation}
where
\begin{align}
  \label{eq:16}
  q(t,\tau)&=\int_{0}^tR(\tau,\tau')f_1(\tau')\,d\tau'\\
  &=\frac{1}{f_1(\tau)}\int_0^t\left[f_2(\tau,\tau')-f_1(\tau)f_1(\tau')\right]\,d\tau'\nonumber
\end{align}
is a measure of upcrossing correlations on the time scale $t$.
The formula Eq.~\eqref{eq:22} has been termed {\it Stratonovich approximation}
\cite{VSS2006b}. Comparing  the Stratonovich approximation with the first-order decoupling approximation, Eq.~\eqref{eq:58}, we observe that the upcrossing rate $f_1(\tau)$ is multiplied by a
correction factor $\ln(1+q)/q$. However, this correction factor depends explicitly on time $t$, which precludes a direct interpretation of the integrand in Eq.~\eqref{eq:22} as the hazard rate (but see \cite{MeeAlb21_arxiv} for a hazard rate approximation of the integrand in the time-homogeneous case).
For the Stratonovich approximation to be applicable, one has to require that
\begin{equation}\label{eq:validity-cond}
  q(t,\tau)>-1
\end{equation}
for all $t$ and $\tau$ so as to keep the argument of the logarithm positive 
\cite{VSS2006b}.

In practice, Eq.~\eqref{eq:22} is not useful as a computational tool. A numerical evaluation is highly inefficient because Eq.~\eqref{eq:22} contains nested integrals on three levels: for each $\tau$ of the outer integral, the
integral $q(t,\tau)$ needs to be evaluated independently for each time $t$. Furthermore, the numerical integration of $q(t,\tau)$ is itself computationally complex because  $R(\tau,\tau')$ involves a further integration (taking already into account that one of the two integrals in the definition of $f_2$, Eq.~\eqref{eq:f2}, Sec.~\ref{sec:corr-betw-upcr}, can be evaluated analytically \cite{RicSat83,VSS2006b}; we note that $f_2$ can also be expressed in terms of Owen's T function \cite{MeeAlb21_arxiv}). Therefore, we will further simplify Eq.~\eqref{eq:22} by deriving a local approximation of the hazard rate.



\subsection{The auto-correlation function of level crossings for small time lags}
\label{sec:corr-funct-level}

\begin{figure*}[t!]
  \centering
  \includegraphics{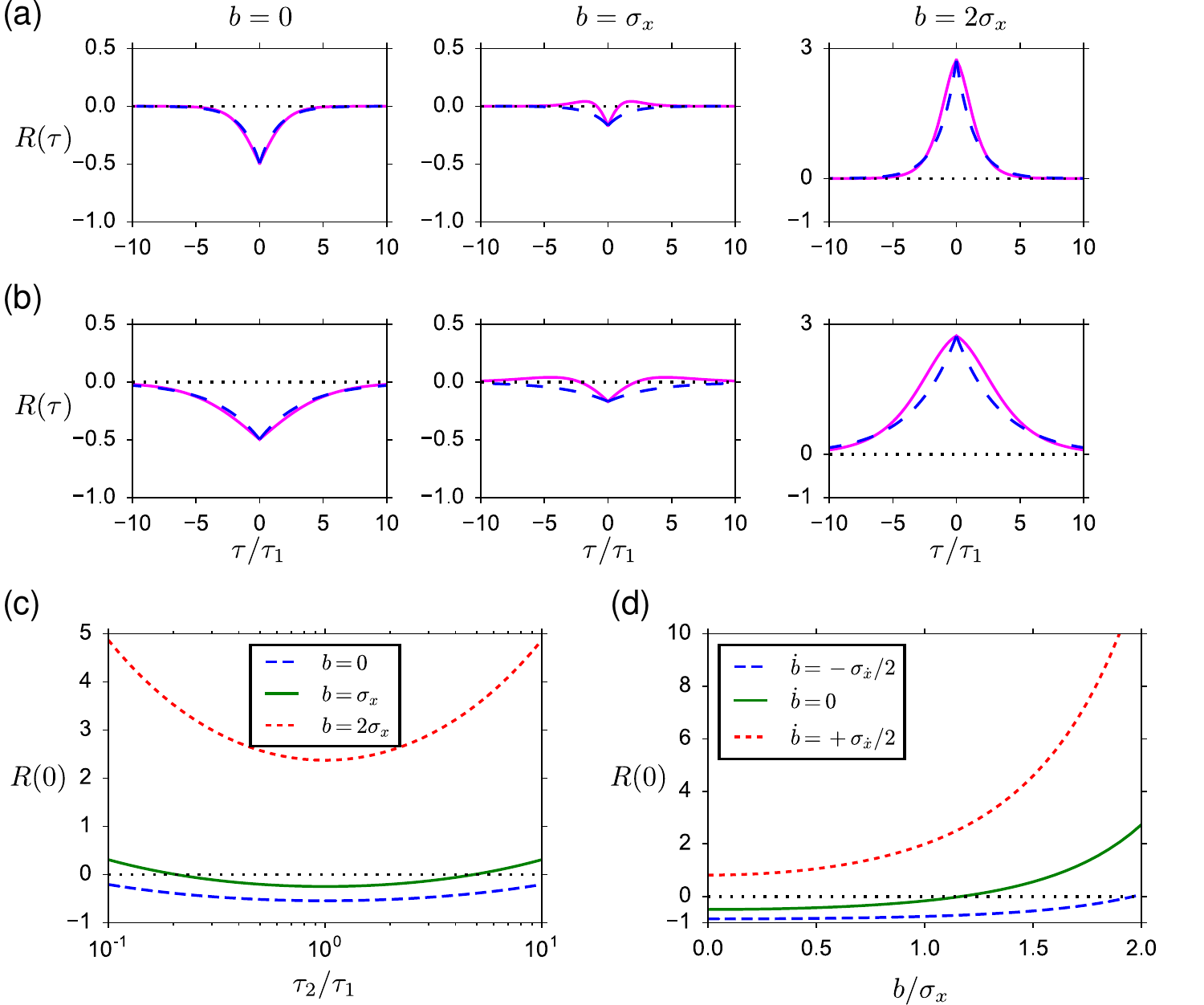}
  \caption{Correlations of level crossings of a stationary process $x(t)$. (a) Normalized auto-correlation function $R(\tau)\equiv R(t,t+\tau)$ as a function on the time lag $\tau$ (in units of $\tau_1\isdef\taum=\gamma^{-1}$, $\tau\neq 0$) for constant barriers $b$ (as indicated on top) and small time constant $\tau_y=0.4\taum$. The solid magenta lines  show the exact semi-analytical result obtained from numerical integration of Eq.~\eqref{eq:f2} and the blue dashed lines shows the exponential approximation, Eq.~\eqref{eq:R-1}, respectively.  (b) Same as (a) but with $\tau_y=2.5\taum$.  (c) Correlations  in the limit of vanishing time lag, $R(0)=\lim_{\tau\rightarrow 0}R(\tau)$, as a function of the time scale ratio $\tau_2/\tau_1=\tau_y/\taum$ for three different constant ($\dot b=0$) threshold levels $b$ (as indicated). (d) Correlations for vanishing time lag as a function of the instantaneous threshold level $b(t)$ for three different slopes $\dot b(t)$ (at $\tau_y=0.4\taum$): decreasing thresholds lower probability of observing two infinitesimally close level crossings (blue dashed line), whereas increasing threshold increase this probability (finely dashed red line) compared to constant thresholds (solid green line). In all panels, black dotted lines indicate the zero baseline corresponding to a Poisson statistics.}
  \label{fig:levelcross-correlations}
\end{figure*}

We now proceed with  calculating the auto-correlation function $R(t,t+\tau)$  in the limit of small time lags $\tau$. Based on the zero-lag limit we then propose a rough estimation of the temporal correlation structure for $\tau>0$, which will be required for the simplification of the Stratonovich approximation in the next section. While the rate of level-crossings has been studied extensively (e.g. \cite{Ric45,Str67II,VSS2006b,TchMal10}), the calculation of second-order statistics such as the auto-correlation function has not received much attention. To the best of our knowledge, closed-form analytical formulas for the auto-correlation function of non-stationary level crossings have not been published previously. \rot{In the Appendix Sec.~\ref{sec:auto-corr-funct-general}, we also provide formulas for the auto-correlation function of general Gaussian level-crossing processes  in the stationary state (see also \cite{BurLew09} for special cases and \cite{Jun94}} for the related but distinct result for the stationary auto-correlation function of the two-state process triggered by level crossings). 

According to Eq.~\eqref{eq:60}, the auto-correlation function at zero time lag is given by 
\begin{equation}
  \label{eq:R0}
  R_0(t)=\frac{f_2(t,t)}{f_1^2(t)}-1,  
\end{equation}
where $f_2(t,t)\equiv\lim_{\tau\rightarrow 0}f_2(t,t+\tau)$ is defined through the limiting procedure $\tau\rightarrow 0$. This corresponds to the continuous part of the auto-correlation function, i.e. $f_2(t,t)$ excludes the singular self-correlation of points given by $f_1(t)\delta(\tau)$.  The correlations between upcrossing in the limit of vanishing lag can be calculated within a saddle-point approximation (see Appendix, Sec.~\ref{sec:corr-betw-upcr}). The result is
\begin{align}
  \label{eq:f2-zerolag}
  f_2(t,t)&=\frac{3\sqrt{3}-\pi}{36\pi^2}\frac{\sigma_y^2/\tau_y}{\sqrt{\sigma_x^2\sigma_y^2-\sigma_{xy}^2}}e^{-B(b,\dot b,t)}\\
  &=:\hat f_2\lrrundd{b(t),\dot b(t),t}
\end{align}

It is instructive to discuss the stationary case, $\dot b=0$ and $t\rightarrow\infty$, in which the pair correlation function $R(\tau)$ for vanishing time lag $\tau$ obtains the simple form
\begin{equation}
  \label{eq:pair-corr-zero-lag}
R_0=\beta\frac{1+\gamma\tau_y}{\sqrt{\gamma\tau_y}}\exp\lrrund{\frac{b^2}{2\sigma_x^2}}-1
\end{equation}
with the numerical constant $\beta=(3\sqrt{3}-\pi)/9\approx 0.228284$. For any fixed value of $\gamma\tau_y$ this expression becomes minimal at $b=0$ (Fig.~\ref{fig:levelcross-correlations}c, blue dashed line). From this we infer that $R_0$ is always positive if $\gamma\tau_y<0.0583757$ (``white noise regime'') or $\gamma\tau_y>17.1304$ (strong friction or large noise correlation time). In this case, upcrossings tend to cluster. In the wide intermediate range $0.0583757<\gamma\tau_y<17.1304$, the sign of $R_0$ depends on the ratio $|b|/\sigma_x$ of barrier height  to standard deviation of  $x(t)$. For vanishing or low barrier height such that $|b|/\sigma_x$ is below the critical value
\begin{equation}
  \label{eq:bcrit}
\frac{b_{crit}}{\sigma_x}=\sqrt{2\ln\lrrund{\frac{\sqrt{\gamma\tau_y}}{\beta(1+\gamma\tau_y)}}},
\end{equation}
the pair correlation function will be negative at small time lags, i.e. upcrossings tend to repel each other. Closer inspection of Eq.~\eqref{eq:pair-corr-zero-lag} shows that for any barrier height $b$, $R_0$ becomes minimal (i.e. most negative) if $\gamma\tau_y=1$. The absolute achievable minimum is found as $R_0=-0.543431$. Therefore, the value $R_0=-1$ expected for non-approaching points is never realized for level crossings \rot{of a doubly low-pass-filtered white noise such as Eq.~\eqref{eq:lif} and Eq.~\eqref{eq:25} for the membrane potential  and overdamped Brownian particle driven by a one-dimensional Ornstein-Uhlenbeck noise, respectively. This result is} in marked contrast to the assumption of non-approaching level crossings made in previous studies \cite{VSS2006b,PueWol16}.  

On the other hand, for high barriers such that $|b|>b_{crit}$, the pair correlation function is positive at small time lags, implying that upcrossing events tend to cluster. Intuitively, upcrossings are mediated by large fluctuations of $x(t)$ in order to reach the high barrier. Once the barrier is reached, $x(t)$ persists at high values for some period because values of $x(t)$ are positively correlated at short time lags. During this period the probability to cross the barrier for a second time is strongly increased. That is, upcrossings tend to cluster in periods on the order of the correlation time of $x(t)$. This clustering corresponds to a positive pair correlation $R_0$

\subsection{Local hazard function.}
\label{sec:local-hazard-funct}

From the Stratonovich approximation, Eq.~\eqref{eq:22}, we obtain the corresponding hazard rate by
differentiating $-\ln S(t)$ with respect to $t$. Using Eq.~\eqref{eq:16}, the result can be
written as
\begin{equation}
  \label{eq:rho-loc}
  \lambda(t)=f_1(t)\left\{F\bigl(q(t,t)\bigr)+\int_{0}^t f_1(\tau)F'\bigl(q(t,\tau)\bigr)R(t,\tau)\,\mathrm{d}\tau\right\}
\end{equation}
where $F(q)=\ln(1+q)/q$. Because of the integral in Eq.~\eqref{eq:rho-loc}, the hazard rate is still non-local in time. In order to obtain a local approximation, we make two \textit{ad~hoc} approximations. First, Eq.~\eqref{eq:rho-loc} can be considerably simplified if $F'(q(t,\tau))$ only weakly depends
on $\tau$ such that we can pull this function out of the integral. Under this assumption and using
again Eq.~\eqref{eq:16}, the hazard rate reduces to the particularly simple form
\begin{equation}
  \label{eq:lam-simp-strat}
  \lambda(t)=\frac{f_1(t)}{1+q(t)},
\end{equation}
where we used the short-hand notation 
\begin{equation}
  \label{eq:3}
  q(t)\equiv q(t,t)=\int_{0}^tR(t,t')f_1(t')\,dt'.
\end{equation}
\rot{The above ad-hoc approximation seems plausible because the pair-correlation function $R(t,\tau)$ is different from zero only in a region of width $|\tau-t|\sim \tau_{corr}$ around its peak at the integration boundary $\tau=t$, where $\tau_{corr}$ is the correlation time defined in Eq.~\eqref{eq:corr-time} below (Fig.~\ref{fig:levelcross-correlations}a,b).  On this time scale,  $q(t,\tau)$ represents indeed a slowly varying function of $\tau$ since it results from an integration over $R$ (cf. Eq.~\eqref{eq:16}). Note that an alternative approximation has been suggested in \cite{MeeAlb21_arxiv}, which neglects the second term in Eq.~\eqref{eq:rho-loc}.}

The formula Eq.~\eqref{eq:lam-simp-strat} reveals a simple relation between the upcrossing rate and the hazard rate, which is the relevant quantity for the FPT: In the absence of correlations between upcrossings, $q=0$, the two rates are equal, while negative correlations (repulsion of up-crossings) increase the hazard rate and positive correlations (attraction or clustering of up-crossings) decreases the hazard rate compared to the up-crossing rate $f_1$.

Second, to find a local estimation of $q(t)$ we need to turn the integral in Eq.~\eqref{eq:3} into a differential equation for $q$. A simple way to achieve this is to use an exponential approximation for the pair correlation function
\begin{equation}
  \label{eq:R-1}
  R(t,t')\approx R_0(t)\exp\lrrund{-\frac{|t-t'|}{\tau_{corr}}},
\end{equation}
where $R_0(t)=f_2(t,t)/f_1^2(t)-1$ is the limit of vanishing time lag
$\tau\rightarrow 0$. Accordingly, the function $f_2(t,t)$ has to be
understood as the limit $\lim_{\tau\rightarrow 0}f_2(t,t+\tau)$, which
\rot{has been} calculated analytically in the \rot{previous} section. Furthermore,
$\tau_{corr}$ is the typical correlation time with which correlations
between upcrossings decay as function of their temporal distance. As a
rough approximation, this correlation time is given by the
correlation time of the stationary process $x(t)$ itself:
\begin{equation}
  \label{eq:corr-time}
  \tau_{corr}=\int_0^\infty \frac{C_{xx}(\tau)}{C_{xx}(0)}\,\mathrm{d}\tau=\taum+\taus.
\end{equation}
Here, $C_{xx}(\tau)$ is the auto-correlation function of $x(t)$ in the stationary state. In fact, comparison of the exponential approximation with numerical evaluation of the exact quadrature formula of the correlation function confirms our choice of $\tau_{corr}$ and also shows that that the exponential ansatz is reasonable as long as $R_0$ is significantly different from zero (Fig.~\ref{fig:levelcross-correlations} a,b, left and right panels). In the crossover region from negative to positive $R_0$ when the barrier height $b$ is increased, the auto-correlation function has both positive and negative phases that are not captured by an exponential function (Fig.~\ref{fig:levelcross-correlations} a,b, middle panels). However, these deviations are less significant because absolute correlations are small in this case.

 Inserting the exponential ansatz Eq.~\eqref{eq:R-1} into Eq.~\eqref{eq:3}, we can pull $R_0(t)$
in front of the integral and obtain:
\begin{equation}
  \label{eq:q1-loc}
  q(t)\approx R_0(t)z(t),
\end{equation}
where $z(t)=\int_0^t\exp\lreckig{-(t-t')/\tau_{corr}}f_1(t')$ defines a new  auxiliary  variable that satisfies the differential equation
\begin{equation}
  \label{eq:q-dyn2}
  \od{z}{t}=-\frac{1}{\tau_{corr}}z+f_1(t)
\end{equation}
with $z(0)=0$. We note that the slightly different ansatz
$R(t,t')\approx R_0(t')\exp\lrrund{-\frac{t-t'}{\tau_{corr}}}$ yields slightly different equations
with similarly good results. In Sec.~\ref{sec:comp-theory-numer}, we will thus only show the results
for the above ansatz, Eq.~\eqref{eq:R-1}.

We note that in the limit of vanishing correlations between upcrossings, $R_0(t)\equiv 0$, the first-order DA $\lambda(t)\approx f_1(t)$ is recovered from Eq.~\eqref{eq:lam-simp-strat}. Thus, the first-order approximation, Eq.~\eqref{eq:n1}, is expected to be valid if 
\begin{equation}
  \label{eq:63}
  |q(t,t)|\ll 1
\end{equation}
for all $t>0$.

In summary, the local hazard rate in the second-order DA is given by
\begin{equation}
  \label{eq:lam-strat}
  \lambda(t)\approx\Phi_2\bigl(b(t),\dot b(t),z(t),t\bigr)\isdef\frac{\Phi_1\bigl( b(t),\dot b(t),t\bigr)}{1+\hat R_0\lrrundd{b(t),\dot b(t),t}z(t)}.
\end{equation}
Here, $\Phi_1$ is given by Eq.~\eqref{eq:n1} and 
\begin{equation}
  \hat{R}_0\lrrundd{b,\dot b,t}=\frac{\hat{f}_2\lrrundd{b,\dot b,t}}{\lreckig{\Phi_1\lrrundd{b,\dot b,t}}^2}-1
\end{equation}
is the zero-lag correlation between up-crossings, Eq.~\eqref{eq:R0}, where $\hat{f}_2$ is given by Eq.~\eqref{eq:f2-zerolag}. In contrast to the first-order approximation $\Phi_1$, the hazard rate $\Phi_2$  depends on the additional local variable $z$ that obeys
\begin{equation}
  \label{eq:q-dyn3}
  \od{z}{t}=-\frac{1}{\tau_{corr}}z+\Phi_1\lrrundd{b(t),\dot b(t),t},\quad z(0)=0.
\end{equation}
Together with Eq.~\eqref{eq:S-ode}, this ordinary differential equation provides an update rule for the numerical evaluation of the FPT density $P(t)$ forward in time.

\subsection{First-passage-time densities}
\label{sec:harmonic-bound}

\begin{figure*}[t]
  \centering
  \includegraphics[type=pdf,ext=.pdf,read=.pdf]{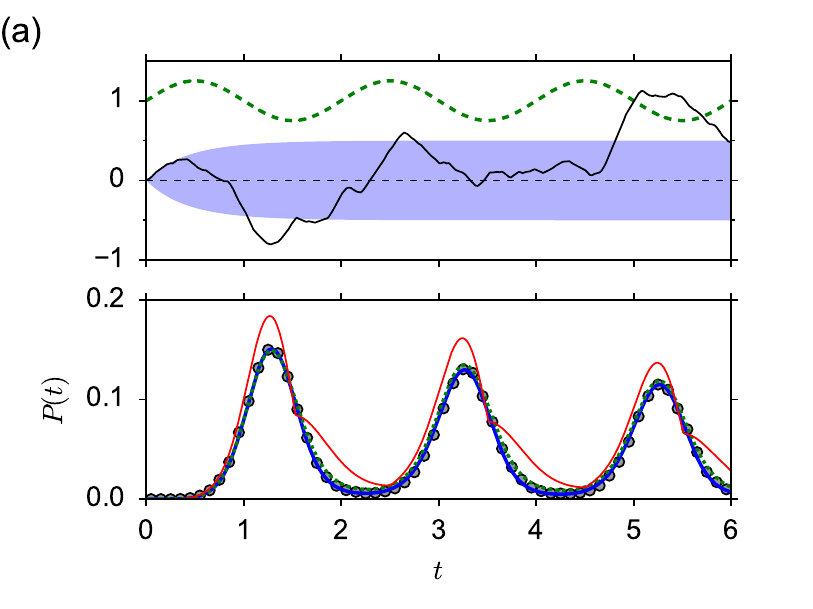}%
  \includegraphics[type=pdf,ext=.pdf,read=.pdf]{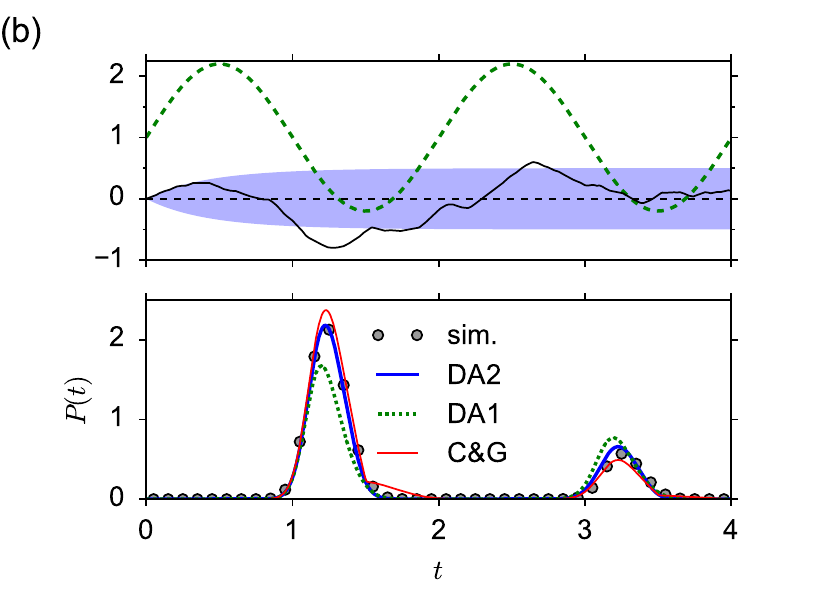}
  \caption{First-passage-time density for periodically moving barrier. (a) Low amplitude $\alpha=0.25$ (subthreshold regime). Top: illustration of moving barrier (green dashed line) and a sample trajectory $x(t)$ (black solid line). The shaded region indicates the mean $\langle x\rangle=0$ (horizontal dashed line) $\pm$ the standard deviation $\sigma_x(t)$. Bottom: First-passage-time density $P(t)$ from simulations (gray circles) and theory (first- and second-order decoupling approximation -- Eq.~\eqref{eq:58} (green dashed line) and Eq.~\eqref{eq:f-lam} (blue solid line), respectively; and the Chizhov-Graham theory -- Eqs.~\eqref{eq:fdrift}-\eqref{eq:lam-chizh} (red thin line)). (b) Same with high amplitude $\alpha=1.2$ (suprathreshold regime). Parameters: $\sigma_x(\infty)=0.5$, $\tau_x=1$, $\tau_y=0.2$, $f=0.5$.}
  \label{fig:fptd-periodic}
\end{figure*}

Being equipped with local approximations of the hazard rate, the FPT density $P(t)$ can be easily obtained from Eq.~\eqref{eq:S-ode}. To test the performance of our theory, we compare the first- and second-order decoupling approximations (DA) with simulations and an alternative hazard-rate theory proposed by Chizhov and Graham \cite{ChiGra08}. An extended variant of the Chizhov-Graham (C\&G) theory is presented in Appendix~\ref{sec:fpt-density-from}, Eq.~\eqref{eq:lam-chizh}.

For concreteness, we consider a periodically moving boundary: 
\begin{equation}
  \label{eq:bounda}
  b(t)=1+\alpha\cos(2\pi ft)
\end{equation}
(Fig.~\ref{fig:fptd-periodic}, top panels). The case, where the amplitude of the oscillating boundary is smaller than unity, $\alpha<1$, corresponds to the subthreshold firing regime of LIF neurons. In this case, both the first- and second-order DA (Eq.~\eqref{eq:S-ode} with $\lambda(t)$ given by Eq.~\eqref{eq:n1} and \eqref{eq:lam-strat}, respectively) yield excellent agreements with simulations (Fig.~\ref{fig:fptd-periodic}a).  In contrast, the C\&G theory (Eq.~\eqref{eq:S-ode} with $\lambda(t)$ given by Eq.~\eqref{eq:lam-chizh}), shows clear deviations from simulations at the peaks of the FPT density and during the time spans when the boundary is increasing ($\dot b>0$), i.e. when the boundary moves away from zero. In these regions, the drift component, Eq.~\eqref{eq:fdrift}, of the C\&G hazard rate is set to zero, leaving only diffusion as a source of threshold crossings. The rectification of the drift component also leads to a characteristic kink at the local extrema of the boundary ($\dot b(t)=0$).

The case of large amplitude oscillations of the boundary ($\alpha>1$) is equivalent to a LIF model that is periodically driven into the supra-threshold regime. In this case, the first-order DA performs significantly  worse than the second-order approximation and the C\&G theory, which both agree well with simulation results (Fig.~\ref{fig:fptd-periodic}b). In particular, the first peak in the FPT density (green dotted line in Fig.~\ref{fig:fptd-periodic}b) is underestimated if correlations between upcrossings are neglected. The underestimation is caused by a reduced hazard rate, which can be understood from the simple formula Eq.~\eqref{eq:lam-simp-strat}: in the first order approximation, the hazard rate is given by the level-crossing rate $\lambda(t)\approx f_1(t)$, while in the second-order approximation $\lambda(t)\approx f_1(t)/[1+q(t)]$ with $q(t)=R_0(t)z(t)$. The factor $1/(1+q)$ accounts for the correlations between upcrossings. At the peak, the boundary $b(t)$ is close to zero.  In this case, the zero lag pair correlation $R_0$ is negative representing the reduced probability of nearby crossings (``repulsion'', Fig.~\ref{fig:levelcross-correlations}, left panels). Since $z$ is positive, we have $-1<q<0$ and thus the factor $1/[1+q]$ is larger than unity (note that $q>-1$ by the assumption Eq.~\eqref{eq:validity-cond}). Therefore, correlations between upcrossings lead to an increased hazard rate and thus a stronger first peak of the FPT density.

\section{Mapping colored input noise to escape noise in the leaky integrate-and-fire model}
\label{sec:from-diffusive-noise}

\subsection{Link function}
\label{sec:link-func}

We now come back to our initial motivation to map colored noise in the input to escape noise in the output of a LIF neuron. 
Having derived the hazard rate $\Phi$ for the FPT with moving boundary $b(t)$, it is easy to formulate the link function $\Psi$ in Eq.~\eqref{eq:escape-noise-model} that provides the escape-noise model corresponding to the LIF model with input noise Eq.~\eqref{eq:lif}. To this end, we only need to shift time such that the FPT starts at time $\tl+\tref$ instead of $t=0$, enforce a zero hazard rate during the absolute refractory period, and express the moving threshold $b(t)$ in terms of the mean membrane potential $u(t)$  for $t>\tl+\tref$ using Eq.~\eqref{eq:b-u}. Accordingly, we also replace the temporal derivative of the moving boundary by
\begin{equation}
  \label{eq:bdot}
  \dot b(t)=-\dot u(t)=\frac{u(t)-\mu(t)}{\taum}
\end{equation}
for $t>\tl+\tref$. The last expression shows that, instead of the two functions $u(t)$ and $\dot u(t)$, one can also use the two functions $u(t)$ and $\mu(t)$ if the stimulus $\mu(t)$ is known. 

With these changes, we obtain the link function in the first-order DA as
\begin{equation}
  \label{eq:map1}
 \Psi_1\lrrundd{u,\dot u,\tau}=\theta(\tau-\tref)\Phi_1(\vth-u,-\dot u,\tau-\tref). 
\end{equation}
Here,  $\theta(t)=\mathbb{1}_{t\ge 0}$ is the Heaviside step function and $\Phi_1$ is given by Eq.~\eqref{eq:n1}. Note that in the first-order DA, the link function $\Psi(u,\dot u,z,\tau)=\Psi_1(u,\dot u,\tau)$ does not depend on an auxiliary variable $z$. In contrast, the 2nd-order DA exhibits an additional auxiliary variable $z$. Taking the last spike time and the absolute refractory period into account, its dynamics reads
  \begin{equation}
  \label{eq:z-dyn}  \dot z=-\frac{z}{\taum+\taus}+\Psi_1(u,-\dot u,t-\tl)
\end{equation}
with initial condition $z(\tl)=0$. We can now write the link function $\Psi$ in the second-order DA as
\begin{equation}
  \label{eq:map2}
  \Psi_2\lrrundd{u,\dot u,z,\tau}=\theta(\tau-\tref)\Phi_2(\vth-u,-\dot u,z,\tau-\tref),
\end{equation}
where $\Phi_2$ is given by Eq.~\eqref{eq:lam-strat}.

\subsection{Comparison with simulation and C\&G theory}
\label{sec:neuron}

\begin{figure*}[t]
  \centering
  \includegraphics[width=7in]{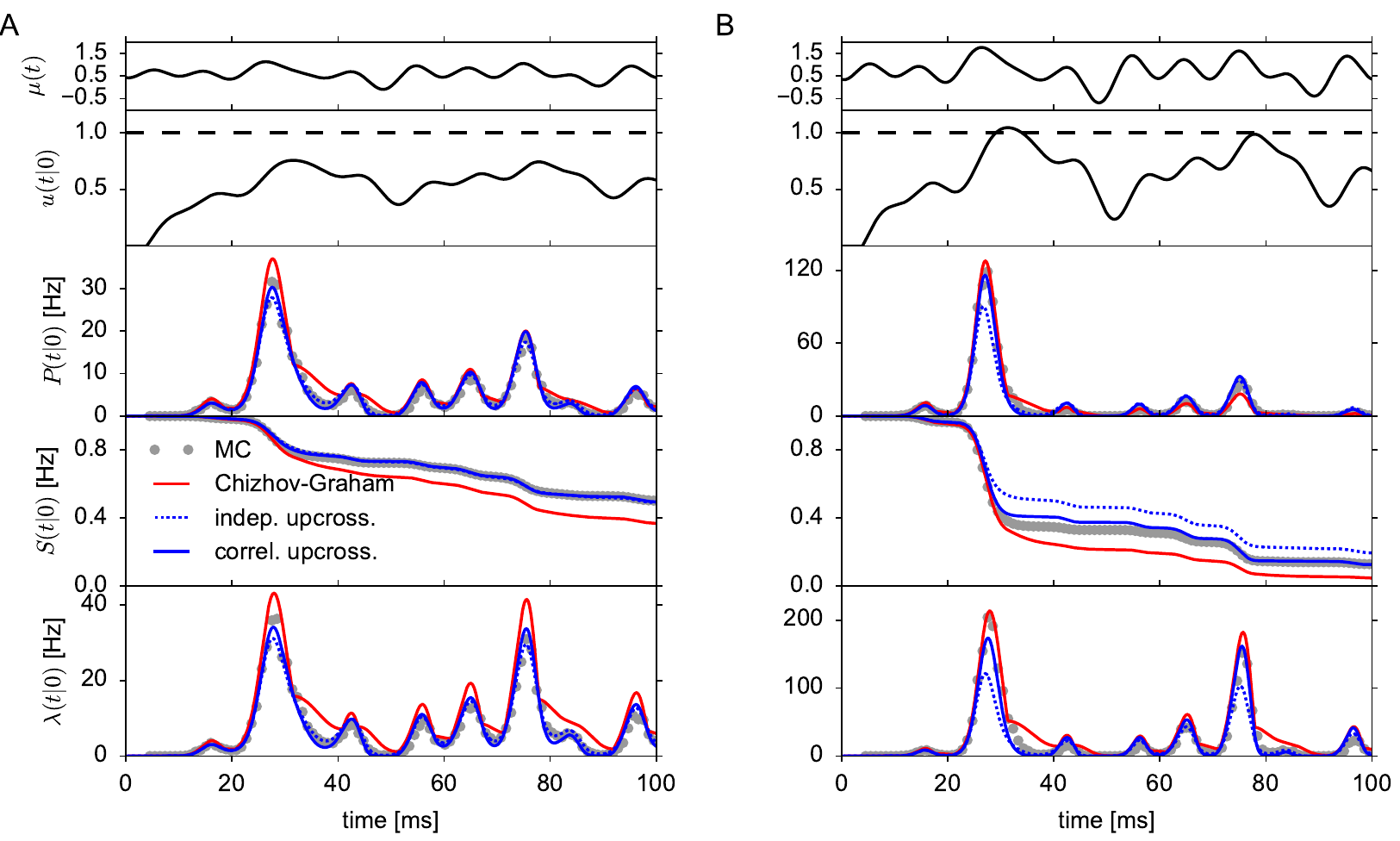}
  \caption{First-passage-time density, survivor function and hazard
    rate under non-stationary driving of a neuron
    that fired its last spike at time $\tl=0$. (A) Weak subthreshold
    stimulus $\mu(t)$ (top panel) leads to a mean membrane potential
    response $u(t|0)$ below threshold at $\vth=1$ (second
    panel). The first-passage-time density $P(t|0)$ for the first
    threshold crossing of $\hat V(t)$ is shown in the third panel
    (gray circles: MC simulations of $10^6$ trials; red solid line:
    Chizhov-Graham theory, Eq.~\eqref{eq:f-lam},~\eqref{eq:lam-chizh}; blue dashed line: first-order decoupling approximation (independent up-crossings), Eq.~\eqref{eq:hazard-1st-ord},~\eqref{eq:isi-gen}; blue
    solid line: second-order decoupling approximation (correlated upcrossings),
    Eq.~\eqref{eq:hazard-2nd-ord},~\eqref{eq:isi-gen}. The
    survival probability $S(t|0)=-dP(t|0)/dt$ and the corresponding
    hazard rate $\lambda(t|0)$ are shown in the two bottom panels,
    respectively. For MC simulations, the hazard rate is computed from
    the ratio $\lambda(t|0)=P(t|0)/S(t|0)$. (B) The same for a
    suprathreshold stimulus, for which the mean membrane potential
    $u(t|0)$ reaches the threshold. In both figures, $\taus=4$~ms, $\taum=10$~ms and $\sigma_\eta$ is such that the standard deviation of $\hat V$ is $\sigma_V=0.25$.}
  \label{fig:fptd-macro}
\end{figure*}

To judge the performance of the level-crossing theory given by the link functions $\Psi_1$ and $\Psi_2$, we compared ISI densities, survival functions and hazard rates with Monte-Carlo simulations of the LIF model with colored input noise, Eq.~\eqref{eq:lif}, and the C\&G theory. These functions are obtained from the link functions as
\begin{align}
  \label{eq:isi-gen}
  \isi(t|\tl)&=\lambda(t|\tl)S(t|\tl),\\
  S(t|\tl)&=\exp\lrrund{-\int_{\tl}^t\lambda(s|\tl)\,ds},
\end{align}
where for the first-order decoupling approximation(DA)
\begin{equation}
  \label{eq:hazard-1st-ord}
  \lambda(t|\tl)\approx\Psi_1\lrrundd{u(t|\tl),\dot u(t|\tl),t-\tl},
\end{equation}
and for the second-order DA, 
\begin{align}
  \label{eq:hazard-2nd-ord}
  \lambda(t|\tl)&\approx\Psi_2\lrrundd{u(t|\tl),\dot u(t|\tl),z(t|\tl),t-\tl}
\end{align}
with $\Psi_1$ and $\Psi_2$ given by Eq.~\eqref{eq:map1} and Eq.~\eqref{eq:map2}, respectively. In Eq.~\eqref{eq:hazard-1st-ord} and \eqref{eq:hazard-2nd-ord}, we have introduced the membrane potential and the auxiliary variable as deterministic functions of $t$ and $\tl$. For $t>\tl+\tref$, these functions obey the first-order dynamics 
\begin{align}
  \dot u(t|\tl)&=-\frac{u(t|\tl)+\mu(t)}{\taum},\\
  \label{eq:z-dyn2}
  \dot z(t|\tl)&=-\frac{z(t|\tl)}{\taum+\taus}+\Psi_1\Bigl(u(t|\tl),-\dot u(t|\tl),t-\tl\Bigr).  
\end{align}
with initial conditions $u(\tl+\tref|\tl)=\vreset$ and $z(\tl+\tref|\tl)=0$.

The time-dependent stimulus $\mu(t)$, shown in Fig.~\ref{fig:fptd-macro} (top panels), was obtained as $\mu(t)=\mu_0+\mu_1(t)$, where $\mu_1(t)$ is a fixed realization of  a band-limited white-noise process with a cut-off frequency of $100\text{ Hz}$. Without loss of generality, we also choose the last spike time as the time origin, $\tl=0$. The membrane potential 
$u(t|0)$ is shown in Fig.~\ref{fig:fptd-macro} (second panel from top). Note that in simulations and figures, we measured voltages in units of $\vth-\vreset$ and chose the arbitrary reference potential such that $\vreset=0$,  and hence $\vth=1$.
 For subthreshold stimuli (Fig.~\ref{fig:fptd-macro}A), 
 $u(t|0)<\vth$, both the first- and second-order decoupling approximations agree well with the interval distribution obtained from simulations of the model with colored input noise. As in the case of periodic subthreshold driving (Fig.~\ref{fig:fptd-periodic}a), the C\&G theory exhibits again clear deviations at the peaks of the ISI density and in periods where the slope of the mean membrane potential is negative, $\dot u(t|0)<0$, (Fig.~\ref{fig:fptd-macro}A, middle panel). The overall performance is better visible in the survival function (Fig.~\ref{fig:fptd-macro}A, second panel from bottom), which is related to the cumulative ISI distribution via $S(t|\tl)=1-\int_{\tl}^tP(s|\tl)\,ds$. It confirms the excellent performance of both decoupling approximations in the subthreshold regime. For completeness, we also compared the hazard rates  (Fig.~\ref{fig:fptd-macro}A, bottom panel).
Note that the initial transient of $u(t|0)$ from reset to resting potential $\mu_0$ realizes a relative refractory period, where the the probability to fire is low.

For suprathreshold stimuli, where the mean membrane potential exceeds the threshold, the first-order DA deviates significantly from simulation results
(Fig.~\ref{fig:fptd-macro}B). This is because level crossings occur more frequently when $u$ is close to the threshold and thus exhibit stronger (negative) correlations. In this case, the assumption of independent upcrossing is no longer valid. Again, the underestimation of the first peak in the ISI density and the hazard rate (dotted lines in Fig.~\ref{fig:fptd-macro}B, middle and bottom panel) if correlations are neglected can be understood from the simple formula Eq.~\eqref{eq:lam-simp-strat}: under the assumption of independent upcrossings, the hazard rate is given by the level-crossing rate $\lambda(t|0)\approx f_1(t)$, while correlations between upcrossings are accounted for in the second-order approximation as $\lambda(t|0)\approx f_1(t)/[1+R_0(t)z(t|0)]$. We have seen that if $u$ is close to the threshold (corresponding to $b=0$), the zero lag pair correlation $R_0$ is negative representing the reduced probability of nearby crossings (``repulsion'', Fig.~\ref{fig:levelcross-correlations}, left panels). Since $z$ is positive, the factor $1/[1+R_0z]$ is larger than unity (Note that $q\equiv R_0z>-1$ by assumption Eq.~\eqref{eq:validity-cond} for the applicability of the Stratonovich approximation). Therefore, correlations between upcrossings lead to an increased hazard rate (2nd-order DA) as compared to the theory with independent upcrossings (1st-order DA) (blue solid vs. blue dotted line in Fig.~\ref{fig:fptd-macro}B, bottom).

\begin{figure*}[t]
  \centering
  \includegraphics[width=\linewidth]{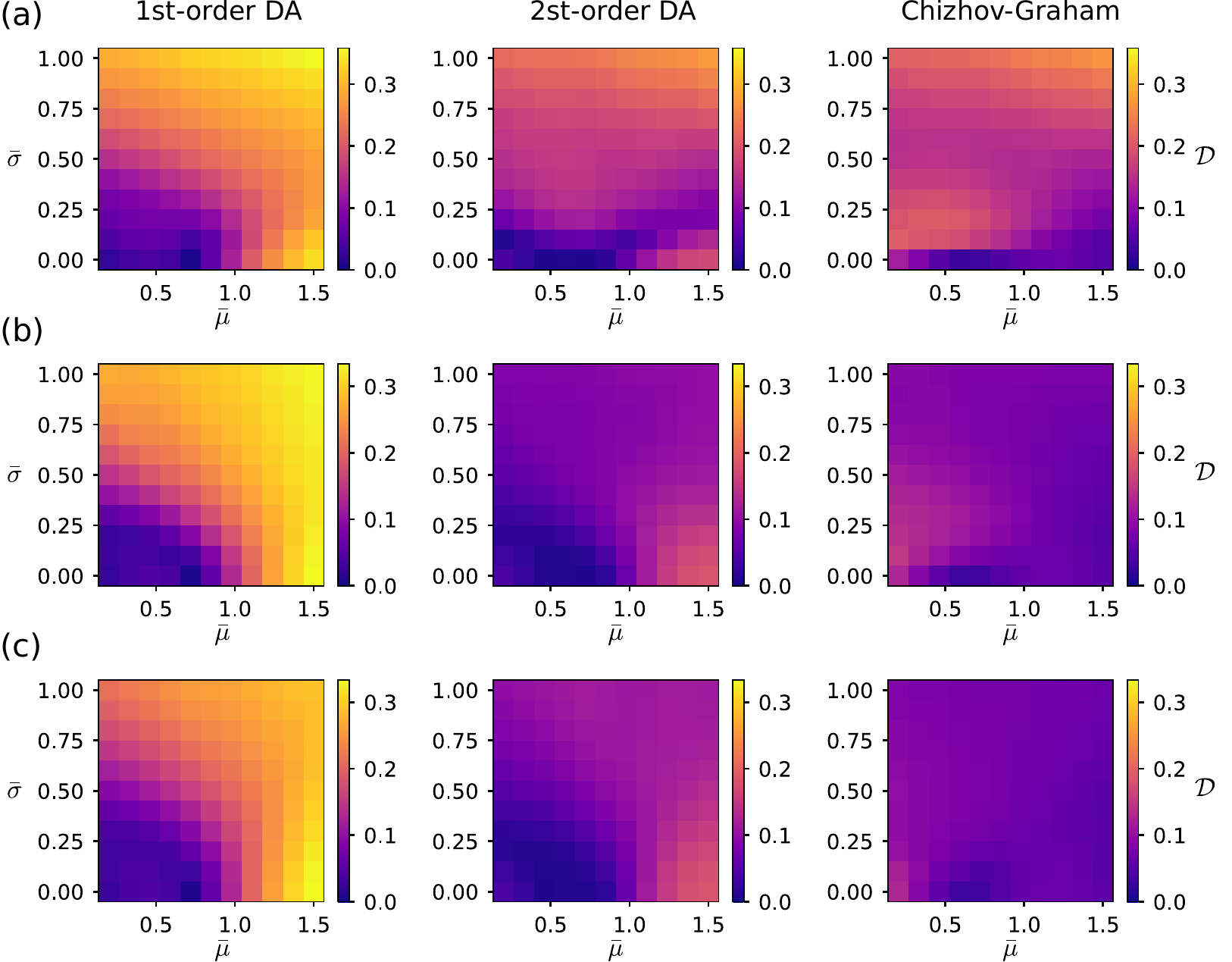}
  \caption{\rot{Error of the theoretical approximations for different stimulus properties. The error is measured as the Kolmogorov-Smirnov distance $\mathcal{D}$ between the theoretical and simulated ISI distribution. The stimulus $\mu(t)$ driving the LIF model is sampled from an Ornstein-Uhlenbeck process with mean $\bar\mu$, standard deviation $\sqrt{1+\taum/\tau_\mu}\bar\sigma$ and correlation time $\tau_\mu$. (a) Color-coded value of $\mathcal{D}$ as a function of $\bar\mu$ and $\bar\sigma$ for a rapidly varying stimulus, $\tau_\mu=1$~ms (left: 1st-order DA , middle: 2nd-order DA, right: Chizhov-Graham theory). (b) Same as (a) but for a moderately fast stimulus, $\tau_\mu=10$~ms. (c) Same as (a) but for a slow stimulus, $\tau_\mu=100$~ms. Other parameters as in Fig.~\ref{fig:fptd-macro}.}}
  \label{fig:K-S}
\end{figure*}

\rot{To characterize the error of the theoretical approximations more systematically, we compare theory and simulations as a function of the stimulus properties (Fig.~\ref{fig:K-S}). To this end, we model $\mu(t)$ as a complex stimulus sampled from a stationary Ornstein-Uhlenbeck process with correlation time $\tau_\mu$, mean $\bar\mu$ and variance $(1+\taum/\tau_\mu)\bar\sigma^2$. This parametrization has been chosen such that the non-resetting membrane potential $\hat V$ has mean $\bar\mu$ and standard deviation $\bar\sigma$ in the stationary state. For a given realization $\mu(t)$, we quantify the deviation of the theoretical ISI distribution $P_\mu(t|0)$ from the simulated one $\hat P_\mu(t|0)$ using the Kolmogorov-Smirnov (KS) statistics \cite{PreTeu92}. This statistics is then averaged over the stimulus ensemble (the subscript $\mu$ indicates the dependence on a given realization $\mu(t)$). Explicitly, the mean KS statistics is defined as
\begin{align}
    \mathcal{D}&=\left\langle\max_{t>0}\left|\int_0^t P_\mu(s|0)\,ds-\int_0^t \hat P_\mu(s|0)\,ds\right|\right\rangle_\mu \nonumber\\
  &=\left\langle\max_{t>0}\left|S_\mu(t|0)-\hat S_\mu(t|0)\right|\right\rangle_\mu,\label{eq:K-S-stat}
\end{align}
where $\langle\cdot\rangle_\mu$ denotes the ensemble average over realizations $\mu(t)$.
Thus, the KS statistics can also be interpreted as the largest absolute difference between the survival function $S_\mu(t|0)$ and the simulated survival function $\hat S_\mu(t)$ (see Fig.~\ref{fig:fptd-macro}, second panels from bottom).

The analysis confirms our previous observations that the decoupling approximations perform best in the subthreshold regime ($\bar\mu<1$) at small stimulus variations $\bar\sigma$ (Fig.~\ref{fig:K-S}); they both become worse in the tonically-firing regime ($\bar\mu>1$). Although the qualitative dependence on the stimulus parameters is similar between the 1st- and 2nd-order DA, the error is considerably smaller for the 2nd-order DA throughout stimulus parameters. On the other hand, the Chizhov-Graham (C\&G) theory has an opposite dependence, it generally performs well in the tonically-firing regime ($\bar\mu>1$) and shows small weaknesses in the subthreshold regime (Fig.~\ref{fig:K-S}b, $\bar\mu<1$), but it exhibits a good overall performance. For all three approximations, the error is larger for a rapidly changing stimulus (Fig.~\ref{fig:K-S}a). Interestingly, in the strongly mean-driven regime ($\bar\mu>1$), a constant or weakly-fluctuating stimulus ($\bar\sigma\ll 1$) turns out to more difficult for the 2nd-order DA than a more strongly fluctuating stimulus (Fig.~\ref{fig:K-S}b,c).}

\section{Population activity of LIF neurons (time-dependent firing rate)}
\label{sec:pop-act}

\subsection{Integral equation}

As an application of the noise mapping, we consider the dynamics of the time-dependent firing rate, or equivalently the population activity of LIF neurons with colored input noise. Being in possession of an approximate hazard rate, it is straightforward to use the renewal integral equation \cite{Ger00,GerKis14} (or equivalently, the refractory density equation \cite{ChiGra07,ChiGra08,DumHen16,SchChi19,PieGal20}) to compute the population activity forward in time. To this end, let us consider a population of $N$ uncoupled LIF neurons with colored input noise, Eq.~\eqref{eq:lif}. The spike train $X_i(t)$ of a given neuron $i$, $i=1,\dotsc,N$ is defined as $X_i(t)=\sum_{k}\delta(t-t_{i,k})$, where $\{t_{i,k}\}_{k\in\mathbb{Z}}$ are the spike times of that neuron. The population activity is defined as the total number of spikes in a small time bin $(t,t+\Delta t)$ divided by the number of neurons $N$ and the time step $\Delta t$. In the limit of infinitely many neurons and infinitesimally small time steps, we obtain the deterministic population activity
\begin{equation}
  \label{eq:pop-act}
  A(t)=\lim_{\Delta t\to\infty}\lim_{N\rightarrow\infty}\frac{1}{N\Delta t}\sum_{i=1}^N\int_t^{t+\Delta t}X_i(t')\,dt'.
\end{equation}
Note that this expression can also be interpreted as an ensemble or trial average of a single neuron spike train, i.e. $A(t)$ is equivalent to the time-dependent firing rate of a single neuron measured over many trials or realizations of a statistical ensemble.  The evolution of the population activity is given by the renewal equation \cite{Cox62,GerKis14}
\begin{equation}
  \label{eq:renewal-eq1}
  A(t)=\isi(t|t_0)+\int_{t_0^+}^t \isi(t|\tl)A(\tl)\,d\tl,
\end{equation}
where $P(t|\tl)$ is given by Eq.~\eqref{eq:isi-gen} and $t_0^+$ denotes the right-sided limit. In Eq.~\eqref{eq:renewal-eq1}, we assumed that the population is initialized with a spike of each neuron at time $t_0$ (``synchronized initial condition''). The first term $P(t|t_0)$ represents the contribution from neurons that fire at time $t$ for the first time after the initial spike at $t_0$.
 The integral equation \eqref{eq:renewal-eq1} can be efficiently solved numerically \cite{GerKis02}. In particular, for numerical solutions, it is useful to turn the exponential factor into a differential equation as in Eq.~\eqref{eq:S-ode}:
\begin{equation}
\od{S(t|\tl)}{t}=-\lambda(t|\tl)S(t|\tl),\qquad S(\tl|\tl)=1  
\end{equation}
for all $\tl<t$.

\begin{figure*}[t]
  \centering
  \includegraphics[width=7in]{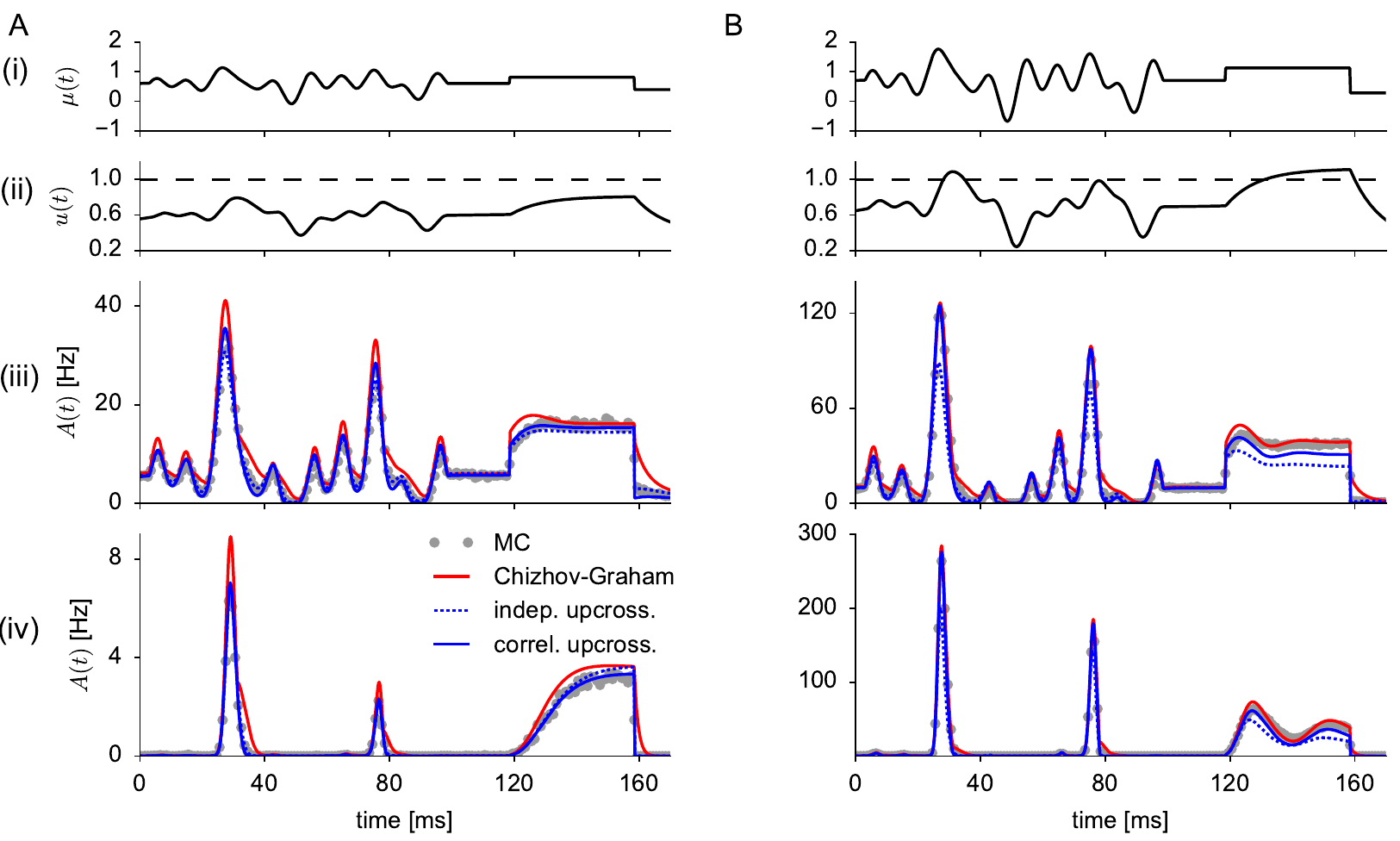}
  \caption{Macroscopic population activity of non-adapting neurons
    under non-stationary driving. (A) Weak subthreshold stimulus
    $\mu(t)$ (i) as in Fig.~\ref{fig:fptd-macro}A leads to a mean
    membrane potential response $u(t|t_0)$ below threshold at
    $\vth=1$ (ii). The resulting population activity $A(t)$ is shown
    in (iii) and (iv) for strong ($\sigma_V=0.25$) and weak
    ($\sigma_V=0.1$) background noise, respectively. Gray circles: MC
    simulations of $10^6$ trials; red solid line: Chizhov-Graham
    theory, Eq.~\eqref{eq:f-lam},~\eqref{eq:lam-chizh}; blue dashed
    line: level-crossing theory with independent upcrossings (1st-order decoupling approximation), Eq.~\eqref{eq:hazard-1st-ord},~\eqref{eq:isi-gen}; blue solid line:
    level-crossing theory with correlated upcrossings (2nd-order decoupling approximation), Eq.~\eqref{eq:hazard-2nd-ord},~\eqref{eq:isi-gen}.  (B)
    The same for a suprathreshold stimulus as in
    Fig.~\ref{fig:fptd-macro}B, for which the mean membrane potential
    $u(t)$ reaches the threshold. In both panels, $\taus=4$~ms, $\taum=10$~ms, $\tref=4$~ms and the population was initialized at time $t_0=-25$~ms (initial transient not shown). }
  \label{fig:rate-macro}
\end{figure*}


\subsection{Comparison with simulations and C\&G theory}
\label{sec:comp-theory-numer}

As an example, we studied the response of the population activity to the complex stimulus $\mu(t)$ shown in Fig.~\ref{fig:rate-macro}Ai and Bi. In the subthreshold regime, where the membrane potential remains below threshold (Fig.~\ref{fig:rate-macro}A), the level-crossing theory well predicts the population activity obtained from simulations, while the C\&G prediction exhibits small deviations as expected from the deviations of the ISI density in the subthreshold regime discussed above (Fig.~\ref{fig:fptd-periodic} and Fig.~\ref{fig:fptd-macro}). The agreement is good for both strong and weak noise.

For suprathreshold stimuli, where the membrane potential exceeds the threshold, the first-order decoupling approximation shows clear deviations (Fig.~\ref{fig:rate-macro}B). However, accounting for correlations between level-crossings in the second-order approximation recovers the population activity of simulated neurons for both strong and weak noise. Similarly, the C\&G theory shows an excellent agreement with simulations.

\section{Discussion}
\label{sec:discussion}

We developed a level-crossing theory for the hazard rate of a leaky integrate-and-fire neuron driven by colored input noise. To this end, we generalized the Stratonovich approximation for the first-passage-time (FPT) density \cite{Str67II,VSS2006b,MeeAlb21_arxiv} to the time-inhomogeneous case, where the stimulus or boundary is time-dependent, and derived a simplification that is local in time. Because higher-order correlations between upcrossings are approximated through their pair-wise correlations, we referred to this theory as the second-order decoupling approximation (DA). Besides the mean membrane potential $u(t)$, the simplified hazard rate depends on the speed $\dot u$ and one additional variable $z(t)$, which accounts for correlations between level crossings. Therefore, the escape-noise model defined by this hazard rate consists of only one extra first-order differential equation, Eq.~\eqref{eq:z-dyn}, besides the dynamics of $u$, Eq.~\eqref{eq:u-dyn}. Our simulation results for the time-dependent interspike-interval (ISI) density and population activity show that the mapped LIF model with escape-noise well matches the LIF model with colored input noise. Thus, the hazard rate  in the 2nd-order DA (link function Eq.~\eqref{eq:map2} and dynamics of $z$, Eq.~\eqref{eq:z-dyn}) provides a novel map from input noise to escape noise. We note that the dependence on the speed $\dot u$ is important and qualitatively differs from commonly used escape-noise models, where the link function only depends on the mean membrane potential $u$. Given the extensive theoretical literature on  population models with simple link functions $\Psi(u)$ \cite{Ger00,CorTan20,SchLoe21_arxiv}, it will be an interesting question for further studies how the mean-field dynamics is influenced by an additional dependence on the membrane potential speed $\dot u$.

The map based on the 2nd-order DA should be compared to the 1st-order DA, which neglects any correlations between upcrossings and represents a time-dependent generalization of the Hertz approximation \cite{VSS2006b}, and the previously proposed map by Chizhov and Graham (C\&G) \cite{ChiGra08}. The generalized Hertz approximation (1st-order DA) involves less ad-hoc approximations compared to the 2nd-order DA (cf. Eqs.~\eqref{eq:lam-simp-strat} and \eqref{eq:R-1}), and performs well in the fluctuation-driven (subthreshold) firing regime at low firing rates. On the other hand, its region of validity, Eq.~\eqref{eq:63}, is rather limited, especially transiently large firing rates and mean-driven (suprathreshold) firing are not well described by the first-order approximation. Furthermore, the gain in numerical efficiency compared to the 2nd-order DA is minor: e.g., simulating the firing rate trajectory of 200ms in Fig.~\ref{fig:rate-macro}B (middle) took 134ms for the 1st-order DA versus 165ms for the 2nd-order DA \rot{(Julia code run on an Intel(R) Core(TM) i7-8550U CPU @ 1.80GHz)}.

On the contrary, the C\&G map exhibits some weaknesses in the fluctuation-driven regime, while it has an excellent performance for short, mean-driven firing-rate transients. This behavior is expected because the theory represents an interpolation between two limit cases, where the theory is expected to work well: strong positive drift towards the threshold without diffusion  (cf. also \cite{GoeDie08}) and pure diffusion without drift. During short mean-driven periods the drift-induced firing dominates and diffusion effects can be safely neglected. An advantage of the C\&G hazard rate, Eq.~\eqref{eq:mapCG}, is its simpler mathematical form and thus easier numerical implementation than the hazard rates based on the level-crossing theory (1st- and 2nd-order DA). Furthermore, the C\&G theory permits to take the white-noise limit, $\taus\to 0$, whereas the level-crossing theory is not well defined in this limit: for $\taus\to 0$, the upcrossing rate $f_1$ diverges \cite{Ric45,Str67II} (cf. Eq.~\eqref{eq:f1-stat-DL}). Despite the divergence in the white-noise limit, we found in simulations that the 2nd-order DA performs well in the physiologically relevant range of synaptic time scales including synaptic time constants as small as $\taus=1$~ms (relative to $\taum=10$~ms, data not shown).
On the other hand, the numerical efficiency of the C\&G and the 2nd-order DA are comparable (e.g. 175ms and  165ms run time, respectively, for the 200ms firing rate trajectory in Fig.~\ref{fig:rate-macro}B, middle). Overall, the C\&G theory represents a good compromise between simplicity and accuracy.

Apart from the mapping of input noise to escape noise, the analysis performed in this paper also provided some analytical insights into the Stratonovich approximation. First, the ansatz of Stratonovich, Eq.~\eqref{eq:DECOUPL-strat}, has been originally proposed for a system of ``non-approaching'' random points (level crossings) \cite{Str67I,Str67II}. In our terminology, this means that the pair correlation function at zero time lag is $R(t,t)=-1$. Put differently, the conditional rate $\nu_{\text{cond}}(t,\tau)=f_2(t,t+\tau)/f_1(t)$ of an upcrossing to occur a time lag $\tau$ after a crossing at time $t$ vanishes for $\tau\to 0$ if upcrossings are non-approaching. However, we found that in our case of the membrane potential driven by an exponentially-correlated Gaussian noise, i.e. a doubly low-pass-filtered white noise (cf. Eq.~\eqref{eq:lif} or \eqref{eq:25}), the upcrossings do not form a system of non-approaching points. The conditional rate $\nu_\text{cond}$ at zero time lag has a non-vanishing minimum (corresponding to a reduced probability of close upcrossings, $\nu_\text{cond}<f_1$) and can even exceed the stationary upcrossing rate, $\nu_\text{cond}>f_1$, (the probability of an upcrossing is increased by an immediately preceding upcrossing, as already noted by \cite{BurLew09} for stationary level-crossings). Given the excellent agreement of the 2nd-order DA with simulations, the ansatz Eq.~\eqref{eq:DECOUPL-strat} seems to be more general and not limited to systems of non-approaching random points.

Based on the assumption of non-approaching level crossings,  the threshold-crossing process has been frequently used as an analytically tractable model of neural spike generation. Examples  include the calculation of information rates \cite{DeWBia}, pairwise correlations and synchronization of neurons
  due to shared inputs \cite{TchMal10,TchGei10,BurLew09} and stochastic
  resonance \cite{Jun95b}. The intuition behind this assumption is that level crossings exhibit refractoriness \cite{PueWol16} or a silence period \cite{TchMal10} because it takes some time for a trajectory to re-cross the threshold from below. While this intuition is true for sufficiently smooth Gaussian processes  \cite{TchMal10,TchGei10} (auto-correlation function must be at least four times differentiable at 0), it fails if the velocity of the process is not differentiable (third derivative of auto-correlation at 0 does not exist), as in the present study and in \cite{VSS2006b,TchWol11,Bad11,PueWol16}. Because neurons exhibit some degree of refractoriness, the Gaussian processes of threshold-crossing neurons should be sufficiently smooth to be useful as a spiking neuron model.

By mapping input noise to escape noise we could apply the renewal integral equation to predict the time-dependent population activity of \emph{infinitely} many LIF neurons with colored input noise. This detour via an approximate escape-noise model allowed us to circumvent the direct numerical solution of the two-dimensional Fokker-Planck equation associated with the LIF model Eq.~\eqref{eq:lif}. An intriguing question is whether the same indirect approach could be used to solve the important problem of \emph{finitely} many neurons with input noise. Neural circuits in the brain are often modeled by networks of integrate-and-fire neurons driven by Poissonian input noise (e.g. \cite{DieGew99,PotDie14,DonSch18}). In these network models, the number of neurons per cell type range from about hundred to a few thousand cells, consistent with experimental estimations \cite{LefTom09}. On this mesoscopic scale, finite-size fluctuations of the population activity cannot be neglected.
It is, however, unknown how to generalize the Fokker-Planck equation to \rot{a stochastic population equation} in the case of finite neuron numbers, so as to account for finite-size fluctuations. On the other hand, the problem of finite-size neural population equations has been recently solved for LIF neurons with escape noise in the form of a stochastic integral equation \cite{SchDeg17,SchLoe21_arxiv}. In the original paper \cite{SchDeg17}, we have applied the stochastic integral equation to the cortical microcircuit model of \cite{PotDie14} by roughly fitting an escape-noise model with the simple link function $\Psi(u)=ce^{\beta u}$ to match mean population activities of simulation data.  However, with the map derived in this paper, where $\Psi$ depends on $u$ and $\dot u$, it should be possible to directly use the stochastic integral equation as a new mesoscopic population model for finite-size populations of LIF neurons driven by colored input noise.


\appendix


\section{FPT density from level-crossing statistics}

\subsection{General expression for survivor function}
\label{sec:gen-surv}

The sequence of upward crossings of the freely evolving, \rot{non-resetting} membrane
potential across the threshold, or shortly the set of ``upcrossings'',
forms a point process $\{t_1,t_2,\dotsc\}$ in time with
$t_i>0$. Thus, the upcrossing times are defined by $\rot{\hat V}(t_i)=\vth$ and
$\rot{\dot{\hat{V}}}(t_i)>0$. As any point process, the upcrossing times for $t>0$ can be fully characterized by the joint
distribution functions $f_1(t_1)$, $f_2(t_1,t_2)$,
$f_3(t_1,t_2,t_3)$, ... (see, e.g. \cite{Str67I,van92}). These
functions are defined such that 
\begin{equation}
  \label{eq:f-def}
  f_k(t_1,\dotsc,t_k)\mathrm{d}t_1\dotsb\mathrm{d}t_k+\mathcal{O}(\mathrm{d}t)
\end{equation}
is the probability to find an upcrossing in each of the
non-overlapping intervals $[t_1,t_1+\mathrm{d}t_1)$, ...,
$[t_k,t_k+\mathrm{d}t_k)$, with sufficiently small intervals
$\mathrm{d}t_1,\dotsc,\mathrm{d}t_k<\mathrm{d}t$ and non-coinciding
arguments $t_i\neq t_j$ for all $i\neq j$. In the case of coinciding
arguments $t_i=t_j$ for some $i\neq j$, the function $f_k$ is
understood to be its limit value for $t_i\rightarrow t_j$.

For our purpose, it will be more convenient to use the correlation
functions $g_1(t_1)$, $g_2(t_1,t_2)$, $g_3(t_1,t_2,t_3)$,
... (see, e.g. \cite{Str67I,van92}). Similar to the joint distribution
functions $\{f_k\}$, the system of correlation functions $\{g_k\}$
completely characterizes the statistics of the upcrossing times. To
define the correlation functions, we first introduce the generating
functional for the $f_k$ given by
\begin{equation}
  \label{eq:gen-func}
  L[v]\equiv\lrk{\prod_{t_i>0}\lrrund{1+v(t_i)}},
\end{equation}
where $v(t)$ is a test function \cite{Str67I,van92}. It can be shown that expanding the generating functional in powers of $v(t)$ yields
\begin{multline}
  \label{eq:L-exp-v}
  L[v]=1+\sum_{k=1}^\infty \frac{1}{k!}\int_{0}^\infty\!\!\dotsb\!\!\int_{0}^\infty f_k(t_1,\dotsc,t_k)\\
\times v(t_1)\dotsb v(t_k)\,\mathrm{d}t_1\dotsb\mathrm{d}t_k,
\end{multline}
i.e. the functions $f_k$ are the expansion coefficients of the
generating functional. Therefore, the joint distribution functions
$f_k$ can be uniquely generated by functional differentiation of
$L[v]$. In analogy to the cumulants of a random variable that are generated from the logarithm of the moment generating function, the correlation functions $g_k$ can be obtained from $\ln L$ as follows:
\begin{equation}
  \label{eq:g-def}
  g_k(t_1,\dotsc,t_k)=\left.\frac{\delta^k\ln L[v(t)]}{\delta v(t_1)\dotsb\delta v(t_k)}\right|_{v(t)\equiv 0}
\end{equation}
In particular, the first two correlation functions read
\begin{align}
  \label{eq:g-f}
  g_1(t)&=f_1(t),\\
  g_2(t_1,t_2)&=f_2(t_1,t_2)-f_1(t_1)f_1(t_2).
\end{align}

By means of the correlation functions, the survivor function $S(t)$, i.e. the probability for
having no upcrossing in the interval $[0,t)$, can be expressed as Eq.~\eqref{eq:S-full}.

\subsection{Moments and correlation functions of the \rot{Gaussian process}}
\label{sec:momen}

In contrast to the vanishing first moments $\langle x\rangle=\langle y\rangle=0$ and the stationary variance $\sigma_y^2=\langle y^2(t)\rangle$, the second moments $\sigma_x^2(t)=\langle x^2(t)\rangle$ and $\sigma_{xy}(t)=\langle x(t)y(t)\rangle$ are time-dependent. They obey the differential equation \cite{Ris84}
\begin{align}
  \label{eq:sigmadgl}
  \od{(\sigma_x^2)}{t}&=-2\left(\gamma\sigma_x^2-\sigma_{xy}\right),\\
\od{\sigma_{xy}}{t}&=-\tilde{\tau}^{-1}\sigma_{xy}+\sigma_y^2.
\end{align}
with $\sigma_y^2=D/\tau_y$, $\tilde{\tau}^{-1}=\gamma+\tau_y^{-1}$ and $\sigma_x^2(0)=\sigma_{xy}(0)=0$. The explicit solution is
\begin{subequations}
  \label{eq:sigmas}
\begin{align}
  \sigma_{xy}(t)&=\tilde{\tau}\sigma_y^2\left(1-e^{-t/\tilde{\tau}}\right),\\
  \sigma_x^2(t)&=\frac{\tilde{\tau}\sigma_y^2}{\gamma}\left(1-e^{-2\gamma t}\right)\nonumber\\
&\quad +\frac{2\tilde{\tau}\sigma_y^2}{2\gamma-\tilde{\tau}^{-1}}\left(e^{-2\gamma t}-e^{-\frac{t}{\tilde{\tau}}}\right)
\end{align}  
\end{subequations}

For large $t$, the process $[x(t),y(t)]$ becomes stationary with the following constant moments
\begin{equation}
  \label{eq:moments-stat}
  \sigma_x^2=\frac{1}{\gamma}\sigma_{xy}=\frac{\tilde{\tau}}{\gamma}\sigma_y^2.
\end{equation}

\subsection{Joint distribution functions for upcrossings}
\label{sec:fk}

Let us denote the point process of the upcrossings by $\{\hat{t}_i\}_{i=1,2,\dotsc}$. The corresponding spike train can be written as
\begin{align}
  s(t)&=\sum_{i=1}^\infty\delta(t-\hat{t}_i),\nonumber\\
  \label{eq:upc-spiketrain}
            &=\lreckig{\dot x(t)-\dot b(t)}\delta\bigl(x(t)-b(t)\bigr)\theta\bigl(\dot x(t)-\dot b(t)\bigr).
\end{align}
\rot{Note that this equation can be seen as an extension of the Kac-Rice formula \cite{AzaWsc09} to moving boundaries.} The joint distribution function  is defined as
\begin{equation}
  \label{eq:f_k-delta}
  f_k(t_1,\dotsc,t_k)=\lrk{s(t_1)\dotsb s(t_k)}
\end{equation}
(for $t_i\neq t_j$ for $i,j=1,\dotsc,k$, $i\neq j$). Substituting Eq.~\eqref{eq:upc-spiketrain} into Eq.~\eqref{eq:f_k-delta} and taking the average yields
\begin{multline}
  \label{eq:fk}
  f_k(t_1,\dotsc,t_k)=\int_{\dot b_1}^\infty\!\!\dotsb\!\!\int_{\dot b_k}^\infty\mathrm{d}\dot x_1\dotsb\mathrm{d}\dot x_k\,\\
\times \lrrund{\dot x_1-\dot b_1}\dotsb\lrrund{\dot x_k-\dot b_k}p_{2k}^{(x,\dot x)}(b_1,\dotsc,b_k,\dot x_1,\dotsc,\dot x_k),
\end{multline}
where $b_i$ and $\dot b_i$ is short-hand for $b(t_i)$ and $\dot b(t_i)$, respectively. Furthermore, 
$p_{2k}^{(x,\dot x)}(x_1,\dotsc,x_k,\dot x_1,\dotsc,\dot x_k)$ is
the joint probability density for the variables $x_i=x(t_i)$ and
$\dot x_i=\dot x(t_i)$. In our case of the two-dimensional Ornstein-Uhlenbeck process, Eq.~\eqref{eq:25}, $p_{2k}^{(x,\dot x)}$ can be simply expressed by the joint probability density $p_{2k}(x_1,\dotsc,x_k,y_1,\dotsc,y_k)$ of the variables $x_i=x(t_i)$ and
$y_i=y(t_i)$:
\begin{equation}
  \label{eq:prob-trafo}
  p_{2k}^{(x,\dot x)}(b_1,\dotsc,b_k,\dot x_1,\dotsc,\dot x_k)=p_{2k}(b_1,\dotsc,b_k,\gamma b_1+\dot x_1,\dotsc,\gamma b_k+\dot x_k).
\end{equation}
Inserting this expression into Eq.~\eqref{eq:fk} yields
\begin{multline}
  \label{eq:fk-y}
f_k(t_1,\dotsc,t_k)=\int_{0}^\infty\!\!\dotsb\!\!\int_{0}^\infty\mathrm{d}w_1\dotsb\mathrm{d}w_k\,w_1\dotsb w_k\\
\times p_{2k}(b_1,\dotsc,b_k,\gamma b_1+\dot b_1+w_1,\dotsc,\gamma b_k+\dot b_k+w_k),
\end{multline}
where we made the substitution $\dot x_i=\dot b_i+w_i$ with new integration variables $w_i$. We note, however, that for higher-dimensional models, it is generally more convenient to directly compute the density $p_{2k}^{(x,\dot x)}$ and use Eq.~\eqref{eq:fk}. For example, for a $(n+1)$-dimensional Gaussian process $\vec{x}(t)=[x(t),y_1(t),\dotsc,y_n(t)]^T$, this density is determined by the time-dependent correlation functions $\langle x(t)x(t+\tau)\rangle$, $\langle x(t)\dot x(t+\tau)\rangle$, $\langle\dot x(t) x(t+\tau)\rangle$ and $\langle \dot x(t)\dot x(t+\tau)\rangle$, which can be obtained from the time-dependent covariance matrix of $\vec{x}(t)$ in a straightforward manner.



\subsection{Uprossing rate $f_1(t)$}
\label{sec:upross-rate-f_1}

Using the moments $\sigma_x^2(t)$, $\sigma_{xy}(t)$ and $\sigma_y^2$ derived in Sec.~\ref{sec:momen}, the
joint probability density of $x$ and $y$ is given by the bivariate
Gaussian distribution
\begin{equation}
  \label{eq:pxy}
  p_2(x,y,t)=\frac{1}{2\pi\sqrt{|C_2|}}\exp\left(-\frac{\sigma_y^2x^2-2\sigma_{xy}xy+\sigma_x^2y^2}{2|C_2|}\right)
\end{equation}
with $|C_2|=\sigma_x^2\sigma_y^2-\sigma_{xy}^2$.  This allows us to calculate the upcrossing rate $f_1(t)$ from
Eq.~\eqref{eq:fk-y}. The integration can be performed analytically resulting in the formula
Eq.~\eqref{eq:n1}.

\subsection{Correlations between upcrossings for small time lag}
\label{sec:corr-betw-upcr}

Here, we are interested in the probability that two upcrossings occur
very close to each other. More precisely, we want to calculate the
probability density $f_2(t,t+\tau)$ in the limit when the distance
$\tau$ between upcrossings goes to zero. 

\subsubsection{Time-dependent boundary}
\label{sec:time-dep-bound}

To this end, we need the
probability density of the four-dimensional vector
$z=[x(t),x(t+\tau),y(t),y(t+\tau)]^T$, which is given by the multivariate
Gaussian distribution 
\begin{equation}
  \label{eq:p4}
  p_4(z)=\frac{1}{4\pi^2\sqrt{|C_4|}}\exp\lrrund{-\frac{1}{2}z^TC_4^{-1}z}.
\end{equation}
This distribution is determined by the correlation matrix $C_4$ with
elements $\lrrund{C_4}_{ij}=\lrk{z_iz_j}$:
  \begin{equation}
    \label{eq:33}
C_4=
    \begin{pmatrix}
      \sigma_x^2(t)&C_{xx}(t,\tau)&\sigma_{xy}(t)&C_{xy}(t,\tau)\\
      C_{xx}(t,\tau)&\sigma_x^2(t+\tau)&C_{yx}(t,\tau)&\sigma_{xy}(t+\tau)\\
      \sigma_{xy}(t)&C_{yx}(t,\tau)&\sigma_{y}^2&C_{yy}(\tau)\\
      C_{xy}(t,\tau)&\sigma_{xy}(t+\tau)&C_{yy}(\tau)&\sigma_{y}^2
    \end{pmatrix},
  \end{equation}
  where we used the notations $C_{xx}(t,\tau)\equiv \lrk{x(t)x(t+\tau)}$,\\
  $C_{xy}(t,\tau)\equiv \lrk{x(t)y(t+\tau)}$ and
  $C_{yx}(t,\tau)\equiv \lrk{y(t)x(t+\tau)}$. Furthermore, note that
  $\sigma_y^2=\lrk{y^2(t)}$ and
  $C_{yy}(\tau)\equiv \lrk{y(t)y(t+\tau)}$ do not depend on time
  because of the stationarity of $y(t)$. The correlation functions for $\tau>0$ can be computed from the regression theorem \cite{Ris84}:
  \begin{align}
    \label{eq:C-tau}
    C_{xx}(t,\tau)&=G_{xx}(\tau)\sigma_x^2(t)+G_{xy}(\tau)\sigma_{xy}(t),\\
    C_{xy}(t,\tau)&=G_{yx}(\tau)\sigma_x^2(t)+G_{yy}(\tau)\sigma_{xy}(t),\\
    C_{yx}(t,\tau)&=G_{xx}(\tau)\sigma_{xy}(t)+G_{xy}(\tau)\sigma_{y}^2(t),\\
    C_{yy}(\tau)&=G_{yy}(\tau)\sigma_y^2(t),
  \end{align}
where we used the elements of the Green's function
\begin{equation}
  \label{eq:greens-func}
  G(\tau)=
  \begin{pmatrix}
G_{xx}(\tau)&G_{xy}(\tau)\\
0&G_{yy}(\tau)    
  \end{pmatrix}
\end{equation}
of the Ornstein-Uhlenbeck process Eq.~\eqref{eq:25}. Using the negative drift matrix of the Ornstein-Uhlenbeck process $A=\bigl(
\begin{smallmatrix}
  \gamma&-1\\
  0&1/\tau_y
\end{smallmatrix}\bigr)
$, the Green's function is obtained from $G(\tau)=e^{-A\tau}$:
\begin{gather}
  \label{eq:Green}
  G_{xx}(\tau)=e^{-\gamma\tau},\qquad   G_{yy}(\tau)=e^{-\tau/\tau_y},\\
  G_{xy}(\tau)=\frac{\tau_y}{1-\gamma\tau_y}\lrrund{e^{-\gamma\tau}-e^{-\tau/\tau_y}}.
\end{gather}

In Eq.~\eqref{eq:33} we also need the time-shifted moments $\sigma_x^2(t+\tau)$ and $\sigma_{xy}(t+\tau)$. These can be obtained from $\sigma_x^2(t)$ and $\sigma_{xy}(t)$ by propagating Eq.~\eqref{eq:sigmadgl}. This yields
\begin{align}
  \label{eq:sig-t-tau}
  \sigma_{xy}(t+\tau)&=e^{-\frac{\tau}{\tilde{\tau}}}\sigma_{xy}+\tilde{\tau}\lrrund{1-e^{-\frac{\tau}{\tilde{\tau}}}}\sigma_{y}^2\\
\sigma_x^2(t+\tau)&=e^{-\frac{2\tau}{\taum}}\sigma_{x}^2+\frac{\sigma_y^2\tilde{\tau}}{\gamma}\bigl(1-e^{-2\gamma\tau}\bigr)\nonumber\\
&\quad+\frac{2\tilde{\tau}(\sigma_{xy}-\sigma_y^2\tilde{\tau})}{1-2\gamma\tilde{\tau}}\bigl(e^{-2\gamma\tau}-e^{-\tau/\tilde{\tau}}\bigr).
\end{align}

Because we are interested in the limit $\tau\rightarrow 0$, we can
expand the moving threshold at time $t$ to linear order such that
\begin{gather}
  \label{eq:b-lin}
b(t+\tau)=b(t)+\dot b(t)\tau+\mathcal{O}(\tau^2).
\end{gather}
The two-point joint density follows from Eq.~\eqref{eq:fk-y} and \eqref{eq:p4}:
\begin{equation}
  \label{eq:f2}
f_2(t,t+\tau)=\frac{1}{4\pi^2\sqrt{|C_4|}}\iint_{0}^\infty\mathrm{d}w_1\mathrm{d}w_2\,w_1w_2\exp\lrrund{-\frac{1}{2}z^TC_4^{-1}z}
\end{equation}
with $ z=[b,b+\dot b \tau,\gamma b+\dot b+w_1,\gamma (b+\dot b\tau)+\dot b+w_2]^T$. A \rot{straightforward but lengthy} series expansion of the exponent $\tilde B=-\frac{1}{2}z^TC_4^{-1}z$ for small $\tau$ yields
\begin{equation}
  \label{eq:Bfunc--}
  \tilde B=-\frac{w_1^2+w_1w_2+w_2^2}{(\sigma_y^2/\tau_y)\tau}-\tilde B_0(b,\dot b,w_1,w_2)+\mathcal{O}(\tau),
\end{equation}
where
\rot{\begin{multline}
  \label{eq:B0til}
  \tilde{B}_0(b,\dot b,w_1,w_2)=\frac{1}{4\sigma_y^2|C_2|}\left\{(\gamma^2\sigma_x^2-2\gamma\sigma_{xy}+\sigma_y^2)b^2\right.
  \\+2[\gamma(\sigma_{xy}^2+\sigma_x^2\sigma_y^2)-2\sigma_{xy}\sigma_y^2]b(\dot b+w_1)\\
  -2\gamma(\sigma_{xy}^2-\sigma_x^2\sigma_y^2)b(\dot b+w_2)\\
  +[(1-\gamma\tau_y)\sigma_x^2\sigma_y^2+(1+\gamma\tau_y)\sigma_{xy}^2](\dot b+w_1)^2\\
  \left. -(1+\gamma\tau_y)(\sigma_{xy}^2-\sigma_x^2\sigma_y^2)(\dot b+w_2)^2\right\}
\end{multline}}
is an $\mathcal{O}(1)$ term. The first term of $\tilde{B}$ is of order
$1/\tau$ and has a maximum at the lower integration boundary
$w_1=w_2=0$. Therefore the neighborhood of the point $w_1=w_2=0$ dominates the integral in the limit $\tau\rightarrow 0$. At this point the term $\tilde B_0(b,\dot b,w_1,w_2)$ coincides with $B(b,\dot b)$ in Eq.~\eqref{eq:Bfunc}. Thus we can write
\begin{equation}
  \label{eq:laplace-trick}
   f_2(t,t+\tau)\sim\frac{1}{4\pi^2\sqrt{|C_4|}}e^{-B(b,\dot b)}I(\tau),\qquad \tau\rightarrow 0
\end{equation}
with the Gaussian integral
\begin{align}
  \label{eq:I-tau}
I(\tau)&=\iint_{0}^\infty\mathrm{d}w_1\mathrm{d}w_2\,w_1w_2\exp\lrrund{-\frac{w_1^2+w_1w_2+w_2^2}{(\sigma_y^2/\tau_y)\tau}}\nonumber\\
&=\frac{9-\sqrt{3}\pi}{27}\frac{\sigma_y^4}{\tau_y^2}\tau^2.  
\end{align}
As a last step, we expand the determinant $|C_4|$ to lowest order in $\tau$:
\begin{equation*}
  \label{eq:determ}
|C_4|=\frac{\left(\sigma_x^2\sigma_y^2-\sigma_{xy}^2\right)\sigma_y^4}{3\tau_y^2}\tau^4+\mathcal{O}(\tau^5).
\end{equation*}
Combining all factors yields the two-point upcrossing density in the limit of zero lag given by Eq.~\eqref{eq:f2-zerolag}.

In the stationary case, $\dot b=0$ and $t\rightarrow\infty$, the formula for $f_2(t,t)$ reduces to
\begin{equation}
  \label{eq:statf2}
f_2^{(s)}(t,t)=\frac{(3\sqrt{3}-\pi)\tau_y}{18\pi(1+\gamma\tau_y)}f_1
\end{equation}
with the stationary upcrossing rate
\begin{equation}
  \label{eq:f1-stat-DL}
  f_1=\frac{1}{2\pi}\sqrt{\frac{\gamma}{\tau_y}}\exp\lrrund{-\frac{b^2}{2\sigma_x^2}}.
\end{equation}
This expression results in the pair correlation function Eq.~\eqref{eq:pair-corr-zero-lag}.

\subsubsection{Auto-correlation function of up-crossings for stationary, differentiable Gaussian processes}
\label{sec:auto-corr-funct-general}

In the stationary case, the rate of upcrossings $f_1$ is constant and the second order distribution function as well as the auto-correlation function of $x$ only depend on the time difference, hence $f_2(t,t+\tau)=f_2(\tau)$ and $C_{xx}(t,t+\tau)=C_{xx}(\tau)$. A classical result for the upcrossing rate is \cite{Ric45}
\begin{equation}
  \label{eq:f1-stat}
  f_1=\frac{\sqrt{|C_{xx}''(0)|/\sigma_x^2}}{2\pi}\exp\lrrund{-\frac{b^2}{2\sigma_x^2}}.
\end{equation}
Here, we derive the asymptotic behavior of $f_2(\tau)$ for small time lag $\tau$. To this end, we expand $C_{xx}(\tau)$
\begin{equation}
  \label{eq:corr-func-expan}
  C_{xx}(\tau)=c_0+\sum_{k=2}^\infty\frac{c_k}{k!}|\tau|^k.
\end{equation}
where $c_k=C_{xx}^{(k)}(0)$ denotes the $k$-th right-sided derivative of the correlation function at zero time lag. Here, we have taken into account that the auto-correlation function is an even function.  Furthermore, we have not included the first-order term $c_1|\tau|$ because the derivative  $C_{xx}'(0)=C_{x\dot x}(0)$ must be zero for differentiable processes $x(t)$, i.e. for velocities $\dot x$ with finite variance.  For example, the one-dimensional Ornstein-Uhlenbeck process is excluded because it exhibits a kink in its auto-correlation function $C_{xx}(\tau)\sim e^{-|\tau|/\tau_{cor}}$ at zero lag (i.e. $c_1<0$) implying an infinite variance of the velocity, $\sigma_{\dot x}^2=-c_2=\infty$,  and hence a diverging up-crossing rate, Eq.~\eqref{eq:f1-stat}. This divergence arises for any one-dimensional Langevin dynamics, for which the velocity $\dot x$ exhibits a white noise component, and reflects the fractal nature of Markovian diffusion processes \cite{Jun94}. In the following, we distinguish three cases, all of which have occurred in previous studies: (i) $c_3\neq 0$ corresponding to a kink in the velocity correlation function $C_{\dot x \dot x}(\tau)=C_{xx}''(\tau)$. This case is considered in the present work as well as in previous models \cite{VSS2006b,TchWol11,Bad11,PueWol16}.  (ii) $c_3=0$ and $c_5\neq 0$ corresponding to a kink in the correlation function of the acceleration $\ddot x(t)$, as in \cite{VSS2006b}. And (iii), $c_3=0$ and $c_5=0$ which occurs, e.g., for smooth correlation functions as used in \cite{TchMal10,TchGei10}.

In the first case, $c_3\neq 0$, i.e. when $C_{xx}''(\tau)$ has a kink at zero lag and thus the acceleration $\ddot x$ has infinite variance as in our model Eq.~\eqref{eq:lif}, we find in lowest-order in $\tau$
\begin{equation}
  \label{eq:f2-i}
  f_2(\tau)\sim\frac{3\sqrt{3}-\pi}{18\pi}f_1\left|\frac{c_3}{c_2}\right|,\qquad \tau\rightarrow 0.
\end{equation}
This expression recovers a previous result obtained in \cite{BurLew09}. 
Furthermore, the case $c_3=0$, $c_5\neq 0$, yields the following lowest-order behavior
\begin{equation}
  \label{eq:f2-ii}
  f_2(\tau)\sim \frac{c_5^{3/2}\exp\lrrund{-\frac{b^2}{2(c_0-c_2^2/c_4)}}}{90\sqrt{15}\pi^2\sqrt{|c_2|^3+c_0c_2c_4}}\tau^{\frac{5}{2}},\qquad \tau\rightarrow 0.
\end{equation}
To the best of our knowledge, this expression is a novel result. Finally, the third case $c_3=0$ and $c_5=0$, yields in lowest-order
\begin{equation}
  \label{eq:f2-iii}
  f_2(\tau)\sim \frac{|c_4^2-c_2c_6|\exp\lrrund{-\frac{b^2}{2(c_0-c_2^2/c_4)}}}{1296\pi^2\sqrt{c_2^2-c_0c_4}}\tau^4,\qquad\tau\rightarrow 0,
\end{equation}
which has been reported before \cite{BurLew09}. In the derivation of Eqs.~\eqref{eq:f2-i}--\eqref{eq:f2-iii}, we have used the Gaussian integral
\begin{equation}
  \label{eq:gaussian-int}
  \iint_0^\infty d\dot x_1d\dot x_2\,\dot x_1\dot x_2\exp\lrrund{-\frac{(\dot x_1+\dot x_2)^2}{\beta}}=\frac{\beta^2}{12}.
\end{equation}

\section{Chizhov-Graham theory}
\label{sec:fpt-density-from}

An elegant approximation for the FPT problem has been put forward by
Chizhov and Graham \cite{ChiGra07,ChiGra08}, which we will state here
without proof. The idea is to construct the hazard function from two
limit cases: First, for an excitatory current that is much faster than
the diffusion time, the probability flux across the threshold is
dominated by the deterministic positive drift, whereas the noise can
be treated as frozen. For a monotonic movement of the mean membrane
potential towards the threshold ($\dot u(t)>0$), one can simply shift the Gaussian
probability density along its time-dependent center and calculate the
survival probability as the total probability mass that is below the
threshold at time $t$:
\begin{align}
  \label{eq:Surv-chich-B}
  S_\text{drift}(t)&=\int_{-\infty}^{b(t)}\frac{\mathrm{d}x}{\sqrt{2\pi}\sigma_x(t)}\exp\lrrund{-\frac{x^2(t)}{2\sigma_x^2(t)}},\\
  &=\frac{1}{2}\lreckig{1+\text{erf}\lrrund{\frac{b(t)}{\sqrt{2}\sigma_x(t)}}}
\end{align}
 In contrast, for negative movement of the center of mass,
i.e. downward and away from the threshold, the survival probability is
kept constant. The hazard rate corresponding to the deterministic drift is given by $-\frac{d}{dt}\ln(S_\text{drift})$ resulting in
\begin{equation}
  \label{eq:fdrift}
  \Phi_\text{drift}(b,\dot b,t)=\frac{2}{\sqrt{\pi}}\left[-\dot T\right]_+\frac{\exp(-T^2)}{\text{erfc}(T)}.
\end{equation}
Following \cite{ChiGra07,ChiGra08}, we introduced the dimensionless quantity
\begin{equation}
  \label{eq:mov-thresh}
T(t)=\frac{b(t)}{\sqrt{2}\sigma_x(t)},
\end{equation}
the temporal derivative of which is given by
\begin{equation}
  \label{eq:dTdt}
  \dot T(t)=\frac{1}{\sqrt{2}\sigma_x}\lrrund{\dot b+\frac{b(\sigma_x^2-\taum\sigma_{xy})}{\taum\sigma_x^2}}.
\end{equation}
The moments $\sigma_x^2(t)$ and $\sigma_{xy}(t)$ have been derived above,
Sec.~\ref{sec:momen} and $\taum=1/\gamma$. Note that the second term in Eq.~\eqref{eq:dTdt}
accounts for the non-stationarity of the variance
$\sigma_x^2(t)$. This term is absent in the original formula in
\cite{ChiGra07,ChiGra08}, which assumed stationary
fluctuations with $\sigma_x(t)=\text{const}.$. This version with stationary fluctuations has also been derived in \cite{GoeDie08}.

Second, the effect of diffusion can be captured in the
quasi-stationary limit case of slow driving. In this case, the
survival probability can be calculated analytically, resulting in the corresponding hazard rate
\begin{align}
  \label{eq:fdiff}
  \Phi_\text{diff}(b)&=\Phi_\text{diff}^\text{wn}(T)\left[1-\left(1+\frac{\taum}{\taus}\right)^{-0.71+0.0825(T+3)}\right],\\
  \Phi_\text{diff}^\text{wn}(T)&=\taum^{-1}\exp\bigl(6.1\cdot 10^{-3}-1.12T-0.25T^2\nonumber\\
  &\quad-0.072T^3-0.0117T^4\bigr).
\end{align}
\rot{Here, the numerical coefficients have been fitted to the exact solution \cite{ChiGra08}.} The total hazard rate is simply given by the sum of the two limit cases: 
\begin{equation}
  \label{eq:lam-chizh}
  \lambda(t)=\Phi_\text{drift}(b,\dot b,t)+\Phi_\text{diff}(b,t).
\end{equation}
Thus, we obtain for the hazard rate of the LIF neuron with absolute refractory period $\tref$ and given last spike time $\tl$
\begin{equation}
  \label{eq:hazard-CG}
  \lambda(t|\tl)\approx \Psi_\text{CG}\lrrundd{u(t|\tl),\dot u(t|\tl),t-\tl}
\end{equation}
with the Chizhov-Graham link function
\begin{multline}
  \label{eq:mapCG}
  \Psi_\text{CG}\lrrundd{u,\dot u,\tau}=\theta(\tau-\tref)
  \bigl[\Phi_\text{drift}(\vth-u,-\dot u,\tau-\tref)\\
  +\Phi_\text{diff}(\vth-u,\tau-\tref)\bigr]. 
\end{multline}



\begin{acknowledgements}
I would like to thank Sven Goedeke and Markus Diesmann for numerous inspiring and fruitful discussions throughout this project, especially about applications to cortical synchronization dynamics, Alexander van Meegen for interesting discussions on the Stratonovich approximation and for sharing his unpublished manuscript, and Wulfram Gerstner for his support during part of this project. 
\end{acknowledgements}

%
\section*{Conflict of interest}

The author declares that he has no conflict of interest.

\section*{Code availability}
\label{sec:code}

The code will become available at the following GitHub link after publication:
\begin{verbatim}
https://github.com/schwalger/LIF_hazard_levelcross
\end{verbatim}



\begin{thebibliography}{10}
\providecommand{\url}[1]{{#1}}
\providecommand{\urlprefix}{URL }
\expandafter\ifx\csname urlstyle\endcsname\relax
  \providecommand{\doi}[1]{DOI \discretionary{}{}{}#1}\else
  \providecommand{\doi}{DOI \discretionary{}{}{}\begingroup
  \urlstyle{rm}\Url}\fi

\bibitem{GerKis14}
W.~Gerstner, W.M. Kistler, R.~Naud, L.~Paninski, \emph{Neuronal Dynamics: From
  Single Neurons to Networks and Models of Cognition} (Cambridge University
  Press, Cambridge, 2014)

\bibitem{Kni72}
B.W. Knight, J. Gen. Physiol. \textbf{59}, 734 (1972)

\bibitem{Ger00}
W.~Gerstner, Neural Comput. \textbf{12}, 43 (2000)

\bibitem{AugLad17}
M.~Augustin, J.~Ladenbauer, F.~Baumann, K.~Obermayer, PLoS Comput. Biol.
  \textbf{13}(6), e1005545 (2017).
\newblock \doi{10.1371/journal.pcbi.1005545}

\bibitem{SchDeg17}
T.~{Schwalger}, M.~{Deger}, W.~{Gerstner}, PLoS Comput. Biol. \textbf{13}(4),
  e1005507 (2017).
\newblock \doi{10.1371/journal.pcbi.1005507}

\bibitem{NykTra00}
D.Q. Nykamp, D.~Tranchina, J. Comput. Neurosci. \textbf{8}(1), 19 (2000)

\bibitem{Chi17}
A.V. Chizhov, Biol. Cybern. \textbf{111}(5-6), 353 (2017)

\bibitem{AbbVre93}
L.F. Abbott, C.~van Vreeswijk, Phys. Rev. E \textbf{48}, 1483 (1993)

\bibitem{BruHak99}
N.~Brunel, V.~Hakim, Neural Comput. \textbf{11}, 1621 (1999)

\bibitem{FouBru02}
N.~Fourcaud, N.~Brunel, Neural Comput. \textbf{14}, 2057 (2002)

\bibitem{Ric08}
M.J.E. Richardson, Biol. Cybern. \textbf{99}(4-5), 381 (2008)

\bibitem{Bru00}
N.~Brunel, J. Comput. Neurosci. \textbf{8}, 183 (2000)

\bibitem{PotDie14}
T.C. Potjans, M.~Diesmann, Cereb Cortex \textbf{24}(3), 785 (2014)

\bibitem{SchDro15}
T.~Schwalger, F.~Droste, B.~Lindner, J. Comput. Neurosci. \textbf{39}(1), 29
  (2015).
\newblock \doi{10.1007/s10827-015-0560-x}

\bibitem{LinDoi05}
B.~Lindner, B.~Doiron, A.~Longtin, Phys. Rev. E \textbf{72}(6), 061919 (2005)

\bibitem{Pan04}
L.~Paninski, Netw. Comput. Neural Syst. \textbf{15}(4), 243 (2004)

\bibitem{TruEde05}
W.~Truccolo, U.T. Eden, M.R. Fellows, J.P. Donoghue, E.N. Brown, J.
  Neurophysiol. \textbf{93}(2), 1074 (2005)

\bibitem{PilShl08}
J.W. Pillow, J.~Shlens, L.~Paninski, A.~Sher, A.M. Litke, E.J. Chichilnisky,
  E.P. Simoncelli, Nature \textbf{454}(7207), 995 (2008)

\bibitem{PilLat08}
J.W. Pillow, P.E. Latham, in \emph{Adv. Neural Inf. Process. Syst.} (2008), pp.
  1161--1168

\bibitem{NauGer12}
R.~Naud, W.~Gerstner, PLoS Comput. Biol. \textbf{8}(10) (2012)

\bibitem{BreSen13}
J.~Brea, W.~Senn, J.P. Pfister, J. Neurosci \textbf{33}(23), 9565 (2013)

\bibitem{GalLoe16}
A.~Galves, E.~L\"{o}cherbach, J. Soc. Fr. Stat. \textbf{157}, 17 (2016)

\bibitem{GerDeg17}
F.~Gerhard, M.~Deger, W.~Truccolo, PLOS Computat. Biol. \textbf{13}(2),
  e1005390 (2017)

\bibitem{RaaDit20}
M.B. Raad, S.~Ditlevsen, E.~L{\"o}cherbach, Ann. Inst. H. Poincaré Probab.
  Statist. \textbf{56}(3), 1958 (2020).
\newblock \doi{10.1214/19-AIHP1023}.
\newblock \urlprefix\url{https://doi.org/10.1214/19-AIHP1023}

\bibitem{ChiGra07}
A.V. Chizhov, L.J. Graham, Phys. Rev. E \textbf{75}(1), 011924 (2007)

\bibitem{ChiGra08}
A.V. Chizhov, L.J. Graham, Phys. Rev. E \textbf{77}(1), 011910 (2008)

\bibitem{DumHen16}
G.~Dumont, J.~Henry, C.O. Tarniceriu, J. Theor. Biol. \textbf{406}, 31 (2016).
\newblock \doi{https://doi.org/10.1016/j.jtbi.2016.06.022}

\bibitem{SchChi19}
T.~Schwalger, A.V. Chizhov, Curr. Opin. Neurobiol. \textbf{58}, 155 (2019)

\bibitem{SchGer20}
V.~{Schmutz}, W.~{Gerstner}, T.~{Schwalger}, J. Math. Neurosc. \textbf{10}(5)
  (2020)

\bibitem{SchLoe21_arxiv}
V.~Schmutz, E.~{L\"{o}cherbach}, T.~Schwalger.
\newblock On a finite-size neuronal population equation (2021)

\bibitem{MenNau12}
S.~Mensi, R.~Naud, C.~Pozzorini, M.~Avermann, C.C.H. Petersen, W.~Gerstner, J
  Neurophysiol  (2012)

\bibitem{PozMen15}
C.~Pozzorini, S.~Mensi, O.~Hagens, R.~Naud, C.~Koch, W.~Gerstner, PLoS Comput
  Biol \textbf{11}(6), e1004275 (2015)

\bibitem{TeeIye18}
C.~Teeter, R.~Iyer, V.~Menon, N.~Gouwens, D.~Feng, J.~Berg, A.~Szafer, N.~Cain,
  H.~Zeng, M.~Hawrylycz, et~al., Nat. Commun. \textbf{9}(1), 709 (2018)

\bibitem{ApfLy06}
F.~Apfaltrer, C.~Ly, D.~Tranchina, Netw. Comput. Neural Syst. \textbf{17}(4),
  373 (2006)

\bibitem{BulEls96}
A.~Bulsara, T.C. Elston, C.R. Doering, S.B. Lowen, K.~Lindenberg, Phys. Rev. E
  \textbf{53}, 3958 (1996)

\bibitem{SchTal04}
M.~Schindler, P.~Talkner, P.~Hanggi, Physical Review Letters \textbf{93}(4),
  048102 (2004)

\bibitem{Lin04b}
B.~Lindner, J.~Stat.~Phys. \textbf{117}, 703 (2004)

\bibitem{PleGer2000}
H.E. Plesser, W.~Gerstner, Neural Comput. \textbf{12}, 367 (2000)

\bibitem{HerGer01}
A.~Herrmann, W.~Gerstner, J Comp Neurosci \textbf{11}(2), 135 (2001)

\bibitem{GoeDie08}
S.~Goedeke, M.~Diesmann, New J. Phys. \textbf{10}, 015007 (2008)

\bibitem{VSS2006b}
T.~Verechtchaguina, I.~Sokolov, L.~Schimansky-Geier, Phys. Rev. E \textbf{73},
  031108 (2006)

\bibitem{SchLSG08}
T.~Schwalger, L.~Schimansky-Geier, Phys. Rev. E \textbf{77}, 031914 (2008).
\newblock \doi{10.1103/PhysRevE.77.031914}

\bibitem{SchDie15}
J.~Schuecker, M.~Diesmann, M.~Helias, Phys. Rev. E \textbf{92}, 052119 (2015).
\newblock \doi{10.1103/PhysRevE.92.052119}

\bibitem{Lin04}
B.~Lindner, Phys. Rev. E \textbf{69}, 022901 (2004)

\bibitem{Sch13}
T.~Schwalger, The interspike-interval statistics of non-renewal neuron models.
\newblock Ph.D. thesis, Humboldt-Universit\"{a}t zu Berlin,
  Mathematisch-Naturwissenschaftliche Fakult\"{a}t I (2013).
\newblock
  \urlprefix\url{\url{http://edoc.hu-berlin.de/docviews/abstract.php?id=40328}}

\bibitem{HanTal90}
P.~H\"anggi, P.~Talkner, M.~Borkovec, Rev. Mod. Phys. \textbf{62}, 251 (1990)

\bibitem{Cox62}
D.R. Cox, \emph{Renewal Theory} (Methuen, London, 1962)

\bibitem{Ric45}
S.O. Rice, Bell Syst. Tech. J \textbf{24}, 45 (1945)

\bibitem{RicSat83}
L.M. Ricciardi, S.~Sato, IEEE Trans. Inf. Theory \textbf{29}, 454 (1983)

\bibitem{BraThu17}
W.~Braun, R.~Thul, Phys. Rev. E \textbf{95}, 012114 (2017).
\newblock \doi{10.1103/PhysRevE.95.012114}

\bibitem{AzaWsc09}
J.M. Aza{\"\i}s, M.~Wschebor, \emph{Level sets and extrema of random processes
  and fields} (John Wiley \& Sons, 2009)

\bibitem{Str67I}
R.L. Stratonovich, \emph{Topics in the Theory of Random Noise}, vol.~1 (Gordon
  and Breach, New York, 1967)

\bibitem{van92}
N.G. van Kampen, \emph{Stochastic Processes in Physics and Chemistry}
  (North-Holland, Amsterdam, 1992)

\bibitem{Bad11}
L.~Badel, Phys. Rev. E \textbf{84}, 041919 (2011).
\newblock \doi{10.1103/PhysRevE.84.041919}

\bibitem{Str67II}
R.L. Stratonovich, \emph{Topics in the Theory of Random Noise}, vol.~2 (Gordon
  and Breach, New York, 1967)

\bibitem{PueWol16}
M.~Puelma~Touzel, F.~Wolf, PLOS Comput. Biol. \textbf{11}(12), 1 (2016).
\newblock \doi{10.1371/journal.pcbi.1004636}.
\newblock \urlprefix\url{https://doi.org/10.1371/journal.pcbi.1004636}

\bibitem{MeeAlb21_arxiv}
A.~van Meegen, S.J. van Albada.
\newblock A microscopic theory of intrinsic timescales in spiking neural
  networks (2019)

\bibitem{TchMal10}
T.~Tchumatchenko, A.~Malyshev, T.~Geisel, M.~Volgushev, F.~Wolf, Phys. Rev.
  Lett. \textbf{104}(5), 058102 (2010)

\bibitem{BurLew09}
Y.~Burak, S.~Lewallen, H.~Sompolinsky, Neural Comput. \textbf{21}(8), 2269
  (2009)

\bibitem{Jun94}
P.~Jung, Phys. Rev. E \textbf{50}, 2513 (1994)

\bibitem{PreTeu92}
W.H. Press, S.A. Teukolsky, W.T. Vetterling, B.P. Flannery, \emph{Numerical
  Recipes in C}, 2nd edn. (Cambridge University Press, Cambridge, USA, 1992)

\bibitem{PieGal20}
B.~Pietras, N.~Gallice, T.~Schwalger, Phys. Rev. E \textbf{102}, 022407 (2020).
\newblock \doi{10.1103/PhysRevE.102.022407}

\bibitem{GerKis02}
W.~Gerstner, W.M. Kistler, \emph{Spiking Neuron Models: Single Neurons,
  Populations, Plasticity} (Cambridge University Press, Cambridge, 2002)

\bibitem{CorTan20}
Q.~Cormier, E.~Tanr\'{e}, R.~Veltz, Stoch. Process. Their Appl.
  \textbf{130}(5), 2553 (2020).
\newblock \doi{https://doi.org/10.1016/j.spa.2019.07.010}

\bibitem{DeWBia}
M.~DeWeese, W.~Bialek, Nuovo cimento D \textbf{17}, 733 (1995)

\bibitem{TchGei10}
T.~Tchumatchenko, T.~Geisel, M.~Volgushev, F.~Wolf, Front Comput Neurosci.
  \textbf{4}(1) (2010)

\bibitem{Jun95b}
P.~Jung, Phys. Lett. A \textbf{207}, 93 (1995)

\bibitem{TchWol11}
T.~Tchumatchenko, F.~Wolf, PLOS Comput. Biol. \textbf{7}(10), 1 (2011).
\newblock \doi{10.1371/journal.pcbi.1002239}.
\newblock \urlprefix\url{https://doi.org/10.1371/journal.pcbi.1002239}

\bibitem{DieGew99}
M.~Diesmann, M.O. Gewaltig, A.~Aertsen, Nature \textbf{402}, 529 (1999)

\bibitem{DonSch18}
J.R. Donoso, D.~Schmitz, N.~Maier, R.~Kempter, J. Neurosci. \textbf{38}(12),
  3124 (2018)

\bibitem{LefTom09}
S.~Lefort, C.~Tomm, J.C.F. Sarria, C.C.H. Petersen, Neuron \textbf{61}(2), 301
  (2009)

\bibitem{Ris84}
H.~Risken, \emph{The Fokker-Planck Equation} (Springer, Berlin, 1984)

\end{thebibliography}

\end{document}